\documentclass[11pt]{article}
\usepackage[dvipsnames]{xcolor}
\usepackage[utf8]{inputenc}
\usepackage[margin=2.5cm]{geometry}
\usepackage[english]{babel}
\usepackage{authblk}
\usepackage{abstract}
\usepackage{amsmath,amssymb}
\usepackage{graphicx}
\usepackage{caption}
\usepackage{booktabs}
\usepackage{cite}
\usepackage[hyperfootnotes=true,colorlinks=true,linkcolor=blue,citecolor=purple]{hyperref}
\numberwithin{equation}{section}
\definecolor{purple}{rgb}{0.4,.2,0.7}

\def\ie{\begin{equation}\begin{aligned}}
\def\fe{\end{aligned}\end{equation}}

\begin{document}

\title{\vspace*{-1.5cm}{\normalsize\rm\hfill PUPT-2666}
\\[1.5cm]\LARGE\bf Stringy Corrections from Gravity: Predictions for BFSS and ABJM}

\author[1]{Aidan Herderschee}
\affil[1]{MIT Center for Theoretical Physics -- a Leinweber Institute, Massachusetts Institute of Technology, Cambridge, MA 02139, USA}
\author[2]{Victor Ivo}
\author[2]{Henry W. Lin}
\affil[2]{Joseph Henry Laboratories and Leinweber Forum for Theoretical Physics, Princeton University, Princeton, NJ 08544, USA}
\author[3,4]{Yifan Wang}
\affil[3]{Center for Cosmology and Particle Physics, New York University, New York, NY 10003, USA}
\affil[4]{Institute for Advanced Study, Princeton, NJ 08540, USA}
\date{}

\thispagestyle{empty}

\maketitle

\begin{abstract}
Despite enormous progress in string theory, the leading stringy corrections to many holographic observables remain unknown.
String/M-theory dualities nonetheless allow them to be determined from gravity alone. From the winding modes of one-loop 11d supergravity on a circle, we derive the leading tree-level stringy terms in the type IIA effective action for the metric, dilaton, and Ramond--Ramond one-form, in closed form as a finite supertrace over Lorentz generators. 
On the thermal D0-brane geometry, this gives the first correction to the low-temperature thermodynamics of the BFSS matrix model. Its coefficient differs from published lattice Monte Carlo estimates by $\approx 20\%$. With the four-form flux included, the same method reproduces the $\zeta(3)k^{2}$ term of the ABJM free energy on the three-sphere obtained by localization.
\end{abstract}

\vspace{1in}
\pagebreak
\setcounter{tocdepth}{3}
{\setlength{\parskip}{0pt}\tableofcontents}
\newpage

\section{Introduction}
\label{sec:intro}

String theory is more than supergravity. 
Even at weak coupling ($g_s\to 0$), string theory predicts infinitely many higher-derivative corrections to the Einstein--Hilbert action. Through AdS/CFT~\cite{Maldacena:1997re,Gubser:1998bc,Witten:1998qj}, these corrections control the strong-coupling expansion of the dual field theories. In particular, they are crucial for comparing with lattice and quantum simulations of the dual theories, which are naturally carried out at finite coupling~\cite{Berkowitz:2016jlq,Pateloudis:2022ijr,Rinaldi:2021jbg,Maldacena:2023acv}. Recent work has developed an orbifold lattice framework for simulating $SU(N)$ Yang--Mills theories on fault-tolerant quantum computers, with the Yang--Mills matrix model as an illustrative example~\cite{Halimeh:2024bth}. Yet despite the enormous progress in understanding perturbative and non-perturbative aspects of string theory, the {\it leading} non-linear correction to the Wilsonian effective action has never been fully written down beyond the NS--NS sector~\cite{Paulos:2008tn,Garousi:2020gio,Liu:2022bfg,Liu:2025uqu}. In type IIA supergravity, besides the metric and the dilaton, there are various gauge fields, including the Ramond--Ramond one-form $C_\mu$, and the leading correction at order $\alpha'^3$ completes the well-known $R^4$ term~\cite{Gross:1986iv,Grisaru:1986vi} with terms of up to eight powers of its field strength, schematically $F^8$. For supersymmetric observables, localization~\cite{Pestun:2007rz,Kapustin:2009kz,Pestun:2016zxk} can produce exact answers, which have also been used to constrain higher-derivative couplings in the bulk, such as the $R^4$ and $D^4R^4$ terms~\cite{Chester:2018aca,Binder:2018yvd,Binder:2019mpb,Binder:2019jwn,Chester:2019jas,Chester:2020dja}. In general, however, there is no such shortcut.

With this motivation in mind, we compute the leading $\alpha'$ correction to the tree-level type IIA effective action from a one-loop computation in 11d SUGRA~\cite{Cremmer:1978km}. To do so, we use the duality between M-theory on a circle and type IIA string theory~\cite{Witten:1995ex,Townsend:1995kk}, following Green, Gutperle, and Vanhove~\cite{Green:1997as}. In their one-loop four-graviton amplitude of 11d SUGRA on a circle, the nonzero windings of the loop reproduce the tree-level $R^{4}$ term of type IIA with coefficient $\zeta(3)$~\cite{Gross:1986iv,Grisaru:1986vi}. The zero-winding sector reproduces the one-loop $R^{4}$ term~\cite{Green:1997as,Russo:1997mk,Garousi:2025wfk}. 

Naively, we would trust the eleven-dimensional supergravity calculation
when the M-theory circle is large in Planck units, $R_{11}\gg\ell_p$.
Since $R_{11}/\ell_p=g_s^{2/3}$, this is the strong-coupling regime of
type IIA string theory. The comparison with weakly coupled type IIA is
possible because supersymmetry protects the $R^4$ coupling: its
perturbative expansion contains only the tree-level and one-loop
terms~\cite{Green:1997tv,Green:1997di,Green:1998by,Berkovits:2004px,Wang:2015jna,Wang:2015aua}.
We extend this four-graviton result to the full nonlinear action for the metric, dilaton, and Ramond--Ramond one-form. The explicit computation in this paper applies only to type IIA backgrounds that lift to Ricci-flat 11d metrics with vanishing four-form flux. However, this strategy generalizes to backgrounds with flux.

As a first application, we compute the leading stringy correction to the free energy of the Banks-Fischler-Shenker-Susskind (BFSS) model~\cite{Banks:1996vh}. We work in the regime where the theory admits a type IIA string theory dual~\cite{Maldacena:1997re,Gubser:1998bc,Witten:1998qj,Itzhaki:1998dd}. 
The result agrees with lattice Monte Carlo simulations~\cite{Berkowitz:2016jlq, Pateloudis:2022ijr} to within twenty percent, but is in tension with their published uncertainties. We also apply the computation to Aharony--Bergman--Jafferis--Maldacena (ABJM) theory~\cite{Aharony:2008ug} on the three-sphere, for which the same general method applies, but many simplifications that occur in the BFSS case do not. We find that the ABJM result matches localization calculations in Refs.~\cite{Kapustin:2009kz, Drukker:2010nc, Fuji:2011km, Marino:2011eh, Hanada:2012si}.

\section{Toy model: \texorpdfstring{$\mathcal{N}=1$}{N=1} abelian gauge theory on \texorpdfstring{$\mathbb{R}^{3}\times S^{1}$}{R3 x S1}}
As a toy model of our 11d computation, consider a free 4d $\mathcal{N}=1$ vector multiplet on the background manifold
\begin{equation}
\label{eq:toy_metric}
ds^{2}|_\mathrm{4d}
= h_{ij}(x)\,dx^{i}dx^{j}
   + \rho^{2}(x)\bigl(dz+C_i(x)\,dx^{i}\bigr)^{2}\,,\quad 
z\sim z+L_0\,,\quad i,j=1,2,3\,.
\end{equation}
Our goal is to extract the local two-derivative, nonzero-winding contribution to the 3d effective action for the background metric $h_{ij}$, the Kaluza--Klein (KK) gauge field  (the ``graviphoton'') $C=C_i dx^i$, and the dilaton $\varphi = \frac{1}{2}\log\rho$. This is the familiar setting of effective field theory on a small circle~\cite{DiPietro:2014bca,Benjamin:2023qsc}. All fields (including the fermions) are periodic around the circle, as for the M-theory circle below.\footnote{With antiperiodic fermions, as on a thermal circle, a winding-$w$ fermion loop picks up a sign $(-1)^w$. The boson–fermion cancellation in the supertrace below then fails, and terms with fewer derivatives, such as a cosmological term, survive, as is expected in thermal EFT (see e.g. \cite{Benjamin:2023qsc}).}
For simplicity we restrict the 4d metric to be Ricci-flat (so that the Laplacians are simple for the fields considered here).

We now evaluate the one-loop contribution of winding modes to the effective action. Since the theory is free, the path integral is Gaussian. We integrate over the photon, the Majorana photino, and the Faddeev--Popov ghosts. In Feynman gauge on this Ricci flat background, the effective action is
\begin{equation}
\label{eq:toy_func_trace}
\Gamma = \tfrac{1}{2}\mathrm{Tr}_{V}\log(-\nabla_{V}^{2}) - \tfrac{1}{4}\mathrm{Tr}_{\Sigma}\log(-\nabla_{\Sigma}^{2})   - \mathrm{Tr}_{\mathbf{1}}\log(-\nabla_{\mathbf{1}}^{2})\,,
\end{equation}
Here $R=V,\Sigma,\mathbf{1}$ denotes the vector, spinor,
or scalar representation carried by the photon, photino, or ghosts,
respectively. The covariant derivative is
\begin{equation}
\label{eq:toy_nabla}
\nabla_R=d+\tfrac12\omega_{MN}J_R^{MN},
\qquad
(\nabla_R)_z=\partial_z-\rho A_R,
\end{equation}
where $d$ is the exterior derivative, $\omega_{MN}$ is the spin
connection, $J^{MN}$ are the Lorentz generators in the representation $R$. We normalize them as $(J_{V}^{AB})^{C}{}_{D}=\delta^{AC}\delta^{B}{}_{D}-\delta^{BC}\delta^{A}{}_{D}$ on vectors and $J_{\Sigma}^{AB}=\frac12\Gamma^{AB}$ on spinors. The Riemann tensor is defined by $[\nabla_{M},\nabla_{N}]V^{A}=R^{A}{}_{BMN}V^{B}$ and $R_{MN}=R^{P}{}_{MPN}$, so that the curvature of the connection $\nabla_{R}$ is $\frac12R_{MNAB}J_{R}^{AB}$.
Note that the second equality in \eqref{eq:toy_nabla} holds in a frame invariant under
$K=\partial_z$.

The matrix $A_R$ which encodes the spin connection for fields in representation $R$ along the circle can be written\footnote{Do not confuse the matrix $A_R$ with the Maxwell gauge potential or the KK gauge potential $C$.} as
\begin{equation}
\label{eq:toy_A_KK}
A_R
 = \frac{1}{2\rho}(\nabla_MK_N)J_R^{MN} = (\partial_a\log\rho)J_R^{a\hat z}
    +\frac{\rho}{4}F_{ab}J_R^{ab}\,.
\end{equation}
Here $\rho^2=g_{zz}$, $F=dC$, and
$e^{\hat z}=\rho(dz+C)$. Capital indices label four-dimensional
orthonormal directions, while $a,b=1,2,3$ label orthonormal base
directions.

We are interested in \eqref{eq:toy_func_trace} at the leading order in derivatives of the background fields. The winding sum is an expansion in the circle size $\ell$: after integration, a term with $2n$ derivatives comes with $\ell^{2n-3}$, so the two-derivative terms dominate as $\ell\to0$. Each factor of $A_R$ carries one derivative and one Lorentz generator, while every other ingredient in \eqref{eq:toy_func_trace} at the lowest order in derivatives of the background, namely derivatives of $\rho$ and $A_R$ and curvatures of the base, carries more derivatives than generators. Since the supertrace below vanishes on products of fewer than two generators, only terms built from $A_R$ alone survive at two derivatives. We can therefore treat $A_R$ as if it were constant and the base as flat. The similar counting at eight derivatives underlies the 11d computation of Sec~\ref{sec:11d}.
Diagonalizing the kinetic operators gives the eigenvalues (dropping the subscript on $A_R$ below)
\begin{equation}
\label{eq:toy_eigenvalues}
\vec{k}^{\,2} + (q_{n} + iA)^{2}, \qquad q_{n} = \frac{2\pi n}{\ell}, \qquad \ell = \rho L_{0}.
\end{equation}
Here, $\vec{k}$ is the 3d momentum, $n$ labels KK modes, and $\ell$ is the local size of the circle. 
The integral over $\vec{k}$ and the sum over $n$ perform the trace over spacetime functions, leaving a finite-dimensional trace $\mathrm{tr}_{R}$ over the Lorentz representation $R$. 
We combine the three traces into a single supertrace
\begin{equation}
\label{eq:toy_action_str}
\begin{split}
\mathrm{Str} \equiv  \mathrm{tr}_{V}- \tfrac{1}{2}\mathrm{tr}_{\Sigma} - 2\,\mathrm{tr}_{\mathbf{1}}\,,\quad 
\Gamma= \frac{1}{2}\int d^{4}x\sqrt{g}\,\frac{1}{\ell}\sum_{n\in\mathbb{Z}}\int\frac{d^{3}k}{(2\pi)^{3}} \mathrm{Str}\log\big(\vec{k}^{\,2}+(q_{n}+iA)^{2}\big)\,.
\end{split}
\end{equation}

To isolate the nonzero winding modes, we perform Poisson resummation:
\begin{equation}
\label{eq:poisson_id}
\frac{1}{\ell}\sum_{n\in\mathbb{Z}} f(q_{n} + iA) = \sum_{w\in\mathbb{Z}} e^{w\ell A} \int \frac{dq}{2\pi}\, e^{i q w \ell} f(q)\,.
\end{equation}
Here, $w$ is the winding number, and $q$ is the continuous momentum along the circle. The matrix exponential of $A$ is the Lorentz rotation that a field picks up when it winds $w$ times around the circle. For our purposes, it suffices to keep only nonzero $w$ since the $w=0$ term is proportional to $\ell$, thus subleading in the small $\ell$ limit. Then integrating over $\vec{k}$ and $q$ gives 
\ie 
\label{eq:toy_final}
\Gamma &\supset
\frac{-1}{32 \pi^2}\int d^4x\sqrt g
\sum_{w\neq0}\int_0^\infty\! \!\frac{ds}{s^3}
e^{-(w\ell)^2/(4s)}
\mathrm{Str}\,e^{w\ell A} 
\\
&= -\frac{1}{2\pi^{2}}\int d^{4}x\sqrt{g}
\sum_{w\neq0}
\frac{\mathrm{Str}\,e^{w\ell A}}{(w\ell)^{4}} 
= -\frac{1}{12}\int d^{4}x\sqrt{g}\,
\frac{\mathrm{Str}\,(A^{2})}{\ell^2}
+\cdots \,.
\fe 
Here $s$ can be interpreted as proper time along the worldline of the superparticle winding $w$ times around a circle of length $\ell$.

Since $\mathrm{Str}$ vanishes on products of fewer than two generators, the expansion in powers of $A$ starts at second order.\footnote{Because the bosonic and fermionic degrees of freedom match, $\mathrm{Str}$ vanishes on the identity matrix. Since each Lorentz generator is traceless, $\mathrm{Str}$ also vanishes on a single generator. Thus, all terms with fewer than two generators drop out.} 
Explicitly, for a pair of eigenvalues $\pm i\theta$ of the real antisymmetric matrix $A$, defining $x=\ell\theta$, the sum over $w$ in \eqref{eq:toy_final} can be easily evaluated using properties of polylogarithms,
\begin{equation}
\label{eq:toy_wsum}
\sum_{w\neq0}\frac{\cos(wx)}{w^{4}}=\frac{\pi^{4}}{45}-\frac{\pi^{2}x^{2}}{6}+\frac{\pi|x|^{3}}{6}-\frac{x^{4}}{24}\,,\quad |x|\le2\pi\,.
\end{equation}
The $x^{2}$ term gives the last equality of \eqref{eq:toy_final} and the $x^{4}$ term carries more derivatives. The $|x|^{3}$ term is the one-loop term from the $n=0$ modes (which are massless for vanishing $A$).
 
Defining the dilaton $\rho=e^{2\varphi}$, we can write the effective action explicitly as
\begin{equation}
\label{eq:toy_action_3d}
\begin{split}
\Gamma_2 =\frac{1}{8L_0}\int d^3x\,\sqrt h\,e^{-2\varphi}
\left[
4(\nabla\varphi)^2
+\frac{e^{4\varphi}}{8}F^2
\right].
\end{split}
\end{equation}
This is the two-derivative part of the thermal effective action of a 4d theory on a circle~\cite{DiPietro:2014bca, Benjamin:2023qsc}, with the 4d metric taken to be Ricci-flat. The cosmological term vanishes because the fermions are periodic. The massless $n=0$ modes contribute only the non-analytic $|x|^{3}$ term in \eqref{eq:toy_wsum}, so they do not affect \eqref{eq:toy_action_3d}. In 11d, they are the modes of type IIA supergravity, whose loops are of genus one or higher and do not affect the genus-zero $\zeta(3)$ term.\footnote{In \cite{DiPietro:2014bca}, the leading small-circle (high-temperature) terms are Chern--Simons terms whose coefficients are fixed by the 4d anomalies which do not vanish when appropriate background fields are turned on. The toy model has no pure gravitational anomaly, and we are only turning on the graviphoton (no R-symmetry gauge field), so no such term arises. 11d supergravity has no perturbative anomalies at all, so this subtlety does not arise.}

\section{Full 11d calculation}
\label{sec:11d}
We now consider 11d SUGRA on the background
\ie 
\label{eq:11d_metric}
ds^{2}|_{11d} &= h_{ij}dx^{i}dx^{j}+\rho^{2}(dz+C_{i}dx^{i})^{2}\,,\quad 
z \sim z + L_{0}\,.
\fe
Here, $h_{ij}$, $\rho$, and $C_{i}$ depend only on the ten coordinates $x^{i}$. All fields are periodic in $z$. This supersymmetric spin structure is the one required by the duality between M-theory and type IIA. 

For simplicity, and since this suffices for the BFSS application, we require that the Ricci tensor, the four-form flux $G_{4}$, and the gravitino $\psi_{M}$ all vanish.
We fix the de Donder gauge for the metric and the standard gauges for the three-form and the gravitino~\cite{Siegel:1980jj, Sen:2012kpz, Bhattacharyya:2012ye} (see Appendix~\ref{app:abjm}).
The seven kinetic operators of the graviton, the three-form, the Majorana gravitino, and their ghosts then take the form
\begin{equation}
\label{eq:11d_delta}
\Delta_{R} = -\nabla_{R}^{2} + E, \quad E = -\tfrac{1}{4}R_{MNPQ}J_{R}^{MN}J_{R}^{PQ},
\end{equation}
where $R_{MNPQ}$ is the Riemann tensor. Since the Ricci tensor vanishes, $E$ is zero on $V$, $\mathbf{1}$, and $\Sigma$, as in the toy model.
The one-loop effective action is the functional supertrace
\ie 
\label{eq:11d_func_str}
\Gamma = \tfrac{1}{2}\Big[&\mathrm{Tr}_{\mathrm{Sym}^{2}V} + \mathrm{Tr}_{\Lambda^{3}V} - 2\mathrm{Tr}_{\Lambda^{2}V} + \mathrm{Tr}_{V} - 4\mathrm{Tr}_{\mathbf{1}} - \tfrac{1}{2}\mathrm{Tr}_{V\otimes\Sigma} + \tfrac{3}{2}\mathrm{Tr}_{\Sigma}\Big]\log\Delta\,,
\fe 
and the finite-dimensional supertrace $\mathrm{Str}$ is the same weighted combination of the traces $\mathrm{tr}_{R}$. The weights count the ghosts, including the ghosts for ghosts of the three-form, and the Majorana condition. The gauge fixing and the weights (see Table~\ref{tab:abjm_sectors}) are explained in detail in Appendix~\ref{app:abjm} in the more general setting with flux, which will be needed in Section~\ref{sec:abjm}.

 The supertrace $\mathrm{Str}$ vanishes on any product of fewer than eight Lorentz generators, a reflection of supersymmetry~\cite{Fradkin:1982kf}. 
 The matrix $E$ contains two Lorentz generators and two derivatives, while $A$ contains one of each. By the argument of the toy model, every other term has fewer generators than derivatives. At eight derivatives, they therefore have at most seven generators and vanish inside $\mathrm{Str}$. We can therefore treat $E$ and $A$ as constant in the following analysis of the one-loop determinant.

The eigenvalues of the kinetic operators are those of Eq.~\eqref{eq:toy_eigenvalues} shifted by $E$, with $\vec{k}$ now the 10d momentum along the $x^{i}$. Exchanging the sum over KK modes for a sum over winding modes as before, we find in the form of \eqref{eq:toy_final},
\begin{equation}
\label{eq:11d_winding_sum}
\Gamma = -\frac{1}{2}\int d^{11}x\sqrt{g}\sum_{w\neq 0}\int_{0}^{\infty}\frac{ds}{s}\,\frac{e^{-(w\ell)^{2}/(4s)}}{(4\pi s)^{11/2}}\,\mathrm{Str}\big[e^{w\ell A}e^{-sE}\big],
\end{equation}
We emphasize that this expression is exact only for constant $E$ and $A$. On a general background, it gives the eight-derivative, nonzero-winding contribution of $\Gamma$, by the counting argument above.

As alluded to above, as a consequence of the maximal supersymmetry, the leading contribution at small $\ell$ comes from the eight-derivative order in the background fields. To extract this local action explicitly,  we expand in $E$ and $A$. After the proper-time integral, every eight-derivative term carries $|w\ell|^{-3}$, so the sum over $w\neq0$ gives $2\zeta(3)/\ell^{3}$.\footnote{The term $\mathrm{Str}(E^{a}A^{2b})$ with $a+b=4$ comes with $(-1)^{a}(w\ell)^{2b}/(a!\,(2b)!)$ from the exponentials, and the proper-time integral gives $\Gamma(\tfrac{11}{2}-a)\,(4/(w\ell)^{2})^{11/2-a}$. Together they reproduce the coefficients in \eqref{eq:11d_compact}. For eigenvalues $\pm i\theta$ of $A$ and $x=\ell\theta$, the full sums are $\sum_{w\neq0}\cos(wx)/|w|^{11-2a}=2\,\mathrm{Re}\,\mathrm{Li}_{11-2a}(e^{ix})$. Unlike \eqref{eq:toy_wsum}, they are not polynomials. Besides analytic terms with coefficients $\zeta(11-2a-2j)$, they contain the non-analytic term $x^{10-2a}\log|x|$.} As in Eq.~\eqref{eq:toy_wsum}, the $n=0$ modes, here the massless modes of type IIA supergravity, contribute only non-analytic terms, which first appear at ten derivatives.
Inside the supertrace, $E$ and $A$ commute, since reordering them changes a product of eight generators only by terms with fewer generators. The result takes the compact form
\begin{equation}
\label{eq:11d_compact}
\begin{split}
\Gamma_{8} = -\frac{\zeta(3)}{12288\,\pi^{5}}&\int d^{11}x\sqrt{g}\,\, \ell^{-3} \mathrm{Str}\big[(E-A^{2})^{3}(E-9A^{2})\big].
\end{split}
\end{equation}
Since $E$ and $A$ are built from generators in finite-dimensional representations of the 11d Lorentz group, this supertrace is a finite matrix computation.

To reduce to 10d type IIA string theory, we write the metric functions and the period in Eq.~\eqref{eq:11d_metric} in string-frame variables as
\begin{equation}
\label{eq:iia_variables}
h_{ij} = e^{-2\phi/3}g_{ij}, \qquad \rho = e^{2\phi/3}, \qquad L_{0} = 2\pi\sqrt{\alpha'}.
\end{equation}
Here, $g_{ij}$ is the string-frame metric, $\phi$ is the dilaton, $C_{i}$ is the Ramond--Ramond one-form, which is the KK photon from the 11d circle. The period $z\sim z+L_0$ fixes the freedom to rescale $z$ and makes the mass of a single KK mode equal to the D0-brane mass. Integrating Eq.~\eqref{eq:11d_compact} over $z$ and continuing to Lorentzian signature, we find the leading $\alpha'$ correction to the tree-level action,
\ie 
\label{eq:iia_action}
S_{8} &= \frac{\zeta(3)\,\alpha'^{3}}{2\kappa_{10}^{2}\cdot 384}\int d^{10}x\sqrt{-g}\,e^{-2\phi} \, \mathrm{Str}\big[(\hat{E}-\hat{A}^{2})^{3}(\hat{E}-9\hat{A}^{2})\big]\,, \\
2\kappa_{10}^{2} &= (2\pi)^{7}\alpha'^{4}\,, \quad \hat{E} = e^{-2\phi/3}E\,, \quad \hat{A} = e^{-\phi/3}A\,,
\fe 
where $\kappa_{10}$ is the 10d gravitational coupling.  As in the toy model, the one-loop winding sum is an expansion in the proper length $\ell$. Since $\ell=\rho L_0$ with $\rho=e^{2\phi/3}$, each pair of derivatives beyond eight raises the type IIA genus by one \cite{Green:2006gt}, so \eqref{eq:iia_action} is the only tree-level term from the 11d one-loop amplitude, while tree-level terms with more derivatives come from higher loops in 11d \cite{Green:1999pu}. Since it is derived on Ricci-flat backgrounds, \eqref{eq:iia_action} determines the eight-derivative action only up to terms that vanish on the leading-order equations of motion, which can be changed by field redefinitions. In the following, we use the representative \eqref{eq:iia_action} as written, and evaluate it on the leading-order solutions, where this ambiguity drops out of the free energy.

\section{Leading stringy correction to BFSS free energy}
\label{sec:bfss}
We now apply our result to the free energy of the BFSS model, computing the subleading term in its low-temperature expansion~\cite{Hanada:2008ez}. This BFSS model is the dimensional reduction of 10d super Yang--Mills theory to one dimension and describes $N$ D0-branes~\cite{deWit:1988wri,Witten:1995im,Banks:1996vh}. We consider the large-$N$ limit at finite temperature $T$ in the planar 't Hooft regime~\cite{Itzhaki:1998dd}
\begin{equation}
\label{eq:bfss_regime}
N^{-10/21} \ll T/\lambda^{1/3} \ll 1, \qquad \lambda = g^{2}N.
\end{equation}
Here, $\lambda$ is the 't Hooft coupling, held fixed at large $N$. The theory is strongly coupled in this regime.

We set $\alpha' =1$ from now on. The bulk dual of this regime is the near-horizon geometry of $N$ near-extremal D0-branes~\cite{Itzhaki:1998dd,Biggs:2023sqw},
\begin{equation}
\label{eq:d0_background}
d s^{2} = r^{7/2}h\,d\tau^{2} + \frac{dr^{2}}{r^{7/2}h} + r^{-3/2}d\Omega_{8}^{2}\,, \\
\end{equation}
with 
\ie
h = 1 - \frac{r_{0}^{7}}{r^{7}}\,,\quad e^{\phi} = \frac{(2\pi)^{2}}{d_{0}N}\,r^{-21/4}\,,\quad
C_{1} = -i\,\frac{d_{0}N}{(2\pi)^{2}}\,r^{7}\,d\tau\,,\quad d_{0} = 240\pi^{5}.
\fe 
Here, $d\Omega_{8}^{2}$ is the metric of the unit eight-sphere, and $r_{0}$ is the horizon radius. Smoothness at the horizon fixes the period of $\tau$ and hence the temperature,
\begin{equation}
\label{eq:d0_temperature}
T = \frac{7\,d_{0}^{1/3}}{4\pi}\,r_{0}^{5/2}.
\end{equation}
In the regime of Eq.~\eqref{eq:bfss_regime}, the type IIA supergravity description is valid at the horizon. To see this, note that the horizon radius scales as $T^{2/5}$ by \eqref{eq:d0_temperature}. The curvature of \eqref{eq:d0_background} in string units grows as $r^{3/2}$, so at the horizon it scales as $T^{3/5}$ and is small in the strongly coupled regime. Likewise, the string coupling is small at the horizon in the 't Hooft regime. The leading free energy $F$ and energy $E$ follow from the horizon entropy and the first law of thermodynamics~\cite{Klebanov:1996un},
\begin{equation}
\label{eq:bfss_leading}
\frac{F}{N^{2}\lambda^{1/3}} = -\frac{5}{9}\,a_{0}\,T^{14/5}, \qquad
a_{0} = \frac{2^{21/5}\,3^{12/5}\,5^{2/5}\,\pi^{14/5}}{7^{19/5}} \approx 7.41.
\end{equation}
Lattice Monte Carlo simulations reproduce this result~\cite{Anagnostopoulos:2007fw,Catterall:2008yz,Hanada:2008ez,Hanada:2013rga,Kadoh:2015mka,Filev:2015hia,Hanada:2016zxj,Berkowitz:2016jlq,Pateloudis:2022ijr}.

We now consider the leading $\alpha'$ correction to the free energy. Substituting the background of Eq.~\eqref{eq:d0_background} into Eqs.~\eqref{eq:11d_metric} and \eqref{eq:iia_variables} gives the 11d uplift~\cite{Hyakutake:2013vwa}. A constant complex linear change of coordinates brings it to a product of a flat line and 10d Euclidean Schwarzschild~\cite{Biggs:2023sqw},
\ie 
\label{eq:d0_uplift_flat}
&ds_{11}^{2} = -dt'^{2} + c^{2}\Big(h\,dw^{2} + \frac{dr^{2}}{h} + r^{2}d\Omega_{8}^{2}\Big)\,, \quad K = \partial_{z} = ir_{0}^{-7/2}\big(c^{-2}\partial_{t'} + c^{-3}\partial_{w}\big)\,,\\
&c^{3} = \frac{d_{0}N}{(2\pi)^{2}}\,, \quad t' = cr_{0}^{7/2}\,\tau + ic^{-2}r_{0}^{-7/2}\,z, \quad w = ic^{-3}r_{0}^{-7/2}\,z\,.
\fe 
Here, $z$ has period $2\pi$, $t'$ parametrizes the flat line, and $w$ is the Euclidean time of the Schwarzschild factor. Since this uplift is Ricci-flat with no four-form flux, Eq.~\eqref{eq:iia_action} applies. The $\partial_{w}$ component of $K$ rotates local frames around the tip of the Schwarzschild cigar, so $\hat{A}$ is nonzero. In an orthonormal frame of the Schwarzschild factor, $\hat{E}$ and $\hat{A}$ have the profiles
\begin{equation}
\label{eq:bfss_profiles}
\hat{E} = r_{0}^{7}\,r^{-11/2}\,E_{0}, \qquad \hat{A} = r_{0}^{7/2}\,r^{-11/4}\,A_{0},
\end{equation}
where $E_{0}$ and $A_{0}$ are fixed combinations of the Lorentz generators $J_{R}^{MN}$, given explicitly in the Appendix~\ref{app:BFSS}. Thus, $\hat{E}$ and $\hat{A}^{2}$ share the same radial dependence, and the supertrace over the seven representations of Eq.~\eqref{eq:11d_func_str} gives
\begin{equation}
\label{eq:bfss_str}
\begin{split}
&\mathrm{Str}\big[(\hat{E}-\hat{A}^{2})^{3}(\hat{E}-9\hat{A}^{2})\big] = \Big(\frac{r_{0}}{r}\Big)^{28}r^{6}\,Q\,,
\end{split}
\end{equation}
where $Q$ is the supertrace $
Q = -\frac{1633622445}{512} $
of $E_{0}$ and $A_{0}$ in the seven representations, evaluated explicitly in Appendix~\ref{app:BFSS}. As a result, the radial integrand of Eq.~\eqref{eq:iia_action} falls as $r^{-35/2}$, so its integral converges at large $r$. The region near $r$ of order unity then contributes only at relative order $T^{33/5}$. Evaluating Eq.~\eqref{eq:iia_action} on the uncorrected background gives the first-order shift in the logarithm of the thermal partition function~\cite{Gubser:1998nz}, from which we find
\ie 
\label{eq:bfss_result}
\frac{F}{N^{2}\lambda^{1/3}}= -\frac{5}{9}a_{0}\,T^{14/5} - \frac{5}{18}a_{1}\,T^{23/5} + O(T^{29/5})\,,\quad 
a_{1}= -\frac{1728701}{11}\,\frac{3^{14/5}\,\pi^{8/5}\,\zeta(3)}{2^{38/5}\,5^{1/5}\,7^{23/5}} \approx -12.375.
\fe 
This result is a concrete prediction for BFSS in the large-$N$ limit at subleading order in the strong-coupling expansion. 

Our value of $a_{1}$ is in tension with lattice Monte Carlo fits~\cite{Hanada:2008ez,Hanada:2016zxj,Berkowitz:2016jlq,Pateloudis:2022ijr}, the most recent of which give $-9.90 \pm 0.31$~\cite{Pateloudis:2022ijr} and $-10.0 \pm 0.4$~\cite{Berkowitz:2016jlq}. 
A potential source of the discrepancy is the phase-quenched approximation, in which the simulations drop the phase of the Pfaffian~\cite{Berkowitz:2016jlq,Pateloudis:2022ijr,Batra:2026tob}. Alternatively, the systematic error on $a_1$ may have been underestimated by truncating the fits after the $T^{29/5}$ term. A decisive test requires lower temperatures or the computation of higher-order coefficients. Quantum simulations could offer another route to such low temperatures~\cite{Rinaldi:2021jbg,Maldacena:2023acv}. For the BMN matrix model, Gharibyan et al. proposed adiabatic preparation of low-energy states and a thermofield double state~\cite{Gharibyan:2020bab}.

\section{Leading stringy correction to ABJM free energy}
\label{sec:abjm}
We next compute the genus zero eight-derivative $\alpha'^{3}$ correction to the free energy of the ABJM theory~\cite{Aharony:2008ug}. This model is a 3d $\mathcal{N}=6$ Chern--Simons-matter theory with gauge group $\mathrm{U}(N)\times\mathrm{U}(N)$ at levels $k$ and $-k$ and bifundamental matter. It describes $N$ M2-branes probing an orbifold of $\mathbb{C}^{4}$ by $\mathbb{Z}_{k}$. We consider the regime dual to type IIA string theory~\cite{Aharony:2008ug},
\begin{equation}
\label{eq:abjm_regime}
N^{1/5} \ll k \ll N \ ,
\end{equation}
where the lower bound makes the M-theory circle small, so the string coupling is weak. The upper bound makes the curvature radius large in string units. The holographic dual is then type IIA string theory on the product of $\mathrm{AdS}_{4}$ and $\mathbb{CP}^{3}$~\cite{Aharony:2008ug}.

The leading contribution to the three-sphere free energy is~\cite{Drukker:2010nc, Herzog:2010hf}
\begin{equation}
\label{eq:abjm_leading}
F = \frac{\pi\sqrt{2}}{3}\,k^{1/2}N^{3/2}.
\end{equation}
This result follows from the renormalized on-shell action of Euclidean $\mathrm{AdS}_{4}$~\cite{Emparan:1999pm} and matches supersymmetric localization~\cite{Kapustin:2009kz,Drukker:2010nc,Herzog:2010hf}. To compute the $\alpha'^{3}$ correction, we uplift the type IIA background to eleven dimensions~\cite{Nilsson:1984bj,Aharony:2008ug},
\ie 
\label{eq:abjm_uplift}
ds_{11}^{2} = R^{2}\Big(\frac{1}{4}ds_{\mathrm{AdS}_{4}}^{2} + \frac{1}{k^{2}}(d\psi + k\omega)^{2} + ds_{\mathbb{CP}^{3}}^{2}\Big)\,,\quad 
G_{4}  = \frac{3}{8}R^{3}\,\hat{\epsilon}_{4}\,,\quad 
{R^{6} \over \ell_{p}^{6}} = 2^{5}\pi^{2}Nk = {k^2L^4\over \alpha'^2}\,,
\fe 
where $ds_{\mathrm{AdS}_{4}}^{2}$ and $ds_{\mathbb{CP}^{3}}^{2}$ are the unit-radius metrics, and $\hat{\epsilon}_{4}$ is the volume form of unit-radius $\mathrm{AdS}_{4}$. The coordinate $\psi$ has period $2\pi$, and $\omega$ is a one-form with $d\omega=2J$, where $J$ is the K\"ahler form of the unit-radius Fubini--Study metric $ds_{\mathbb{CP}^{3}}^{2}$. In addition, $R$ is the 11d curvature radius, $L$ is the string-frame radius of $\mathbb{CP}^{3}$, and $\ell_{p}$ is the 11d Planck length, set to unity in the last relation. 
Together, the two relations give $L^{4}=32\pi^{2}\lambda\,\alpha'^{2}$ with the 't~Hooft coupling $\lambda=N/k$, so the $\alpha'$ expansion is an expansion in $\lambda^{-1/2}$. The radius of the M-theory circle is $R/k\propto(N/k^{5})^{1/6}\,\ell_{p}$, which is small in the IIA regime \eqref{eq:abjm_regime}.
The internal space is $S^{7}/\mathbb{Z}_{k}$, a circle bundle over $\mathbb{CP}^{3}$. Unlike the D0-brane uplift in Eq.~\eqref{eq:d0_uplift_flat}, this background is supported by the four-form flux $G_{4}$ and is not Ricci-flat, so Eq.~\eqref{eq:iia_action} does not apply as it stands. Instead, the one-loop computation of the previous sections must be extended to backgrounds with four-form flux. This extension retains the coupled kinetic (fluctuation) operators of the graviton, three-form, and gravitino, and it produces a longer eight-derivative action.

We evaluate this term of the one-loop determinant directly on \eqref{eq:abjm_uplift}. The details are in Appendix~\ref{app:abjm}, which gives the kinetic operators in the presence of flux, the short-time heat kernel on the curved base manifold that generalizes the Gaussian kernel behind \eqref{eq:11d_winding_sum}, and the contribution of each sector of the fluctuating fields (see Table~\ref{tab:abjm_sectors}). Since in the resulting eight-derivative term \eqref{eq:abjm_gamma8} the density scales as $R^{-8}$, the total 11d volume as $R^{11}/k$, and, as in \eqref{eq:11d_compact}, the factor $\ell^{-3}\sim k^{3}/R^{3}$ from the sum over windings, all dependence on $R$ cancels, and we find the bulk contribution to the free energy
\begin{equation}
\label{eq:abjm_result}
\delta F_{\rm bulk} = \frac{\zeta(3)}{8\pi^{2}}\,k^{2}.
\end{equation}
At order $\alpha'^3$, the free energy at fixed $N$ receives the same correction as the grand potential evaluated at the uncorrected potential~\cite{Gautason:2025plx, Bobev:2026gir}. The eight-derivative action also corrects the electric-boundary term. However, at fixed $N$, this correction cancels against the resulting shift of the two-derivative boundary term.

Localization reduces the ABJM $S^{3}$ partition function to a matrix integral~\cite{Kapustin:2009kz}. To all orders in $1/N$ at fixed $k$, this integral is given by the Airy function $\mathrm{Ai}$ \cite{Fuji:2011km,Marino:2011eh},
\begin{equation}\label{eq:abjm_airy}
Z=C^{-1/3}e^{A(k)}\,\mathrm{Ai}\big[C^{-1/3}(N-B)\big],\quad C=\frac{2}{\pi^{2}k}\,,\quad B=\frac{k}{24}+\frac{1}{3k}\,,
\end{equation}
where $A(k)$, which is independent of $N$, captures the so-called constant-map contributions~\cite{Drukker:2010nc,Hanada:2012si}. In the IIA regime of Eq.~\eqref{eq:abjm_regime}, $F=-\log Z$ becomes
\begin{equation}
\label{eq:abjm_localization}
F = \frac{\pi\sqrt{2}}{3}\,k^{1/2}\Big(N - \frac{k}{24}\Big)^{3/2} + \frac{\zeta(3)}{8\pi^{2}}\,k^{2} + \ldots\,,
\end{equation}
up to subleading terms that include logarithms of $N$ and $k$. 
In terms of $\lambda=N/k$, the leading term is of order $N^{2}\lambda^{-1/2}$, and the two corrections in \eqref{eq:abjm_localization} have different origins. The shift of $N$ by $k/24$ is a shift of $\lambda$ to $\lambda-1/24$. It comes from the M2-brane charge induced by curvature through the $C_{3}\wedge X_{8}$ coupling of M-theory~\cite{Bergman:2009zh,Aharony:2009fc}. Although this coupling also has eight derivatives, it is topological: it shifts $N$ in the leading two-derivative result, at relative order $\lambda^{-1}$, and carries no $\zeta(3)$. The $k^{2}$ term instead comes from $A(k)$. With $g_{s}=2\pi i/k$ in the topological-string description, genus $g$ scales as $k^{2-2g}$, so the $k^{2}$ term is the genus-zero constant map~\cite{Drukker:2010nc,Hanada:2012si,BenettiGenolini:2026kjy}. It is planar and suppressed by $\lambda^{-3/2}\propto(\alpha'/L^{2})^{3}$ relative to the leading term, as expected for the genus-zero eight-derivative correction~\cite{Beccaria:2023hhi}. It agrees exactly with the bulk result 
\eqref{eq:abjm_result}.

We also checked \eqref{eq:abjm_result} by directly computing one-loop determinants on ${\rm AdS}_4$, resolving a puzzle in \cite{Liu:2016dau}. Summing the one-loop determinants of the KK tower on $S^{7}/\mathbb{Z}_{k}$ over the continuous spectrum on $\mathrm{AdS}_{4}$ gives terms growing as $\zeta(7)k^{6}$ and $\zeta(5)k^{4}$, and a different $\zeta(3)k^{2}$ coefficient. The normalizable harmonic two-forms on $\mathrm{AdS}_{4}$ carry three-form modes that are not pure gauge. They cancel the first two terms and supply the missing part of the third, reproducing Eq.~\eqref{eq:abjm_result} (see Table~\ref{tab:abjm_spectral} in the Appendix~\ref{app:abjm}).\footnote{This was also independently done in \cite{Kurlyand:2026cyz}.} Evaluating the type IIA eight-derivative couplings, determined directly in 10d in companion work~\cite{Herderschee:2026IIR4}, on $\mathrm{AdS}_{4}\times\mathbb{CP}^{3}$ would give a further check at the level of the action.

A different bulk derivation was recently proposed in 
\cite{BenettiGenolini:2026kjy}, which localizes the perturbative M-theory path integral. There the $\zeta(3)k^{2}$ term arises from a conjectured equivariant one-loop factor valid at finite $k$. Our local action derivation of \eqref{eq:abjm_result} supports that factor at genus zero and, unlike localization, also applies to non-supersymmetric backgrounds such as the thermal D0-brane geometry of Sec~\ref{sec:bfss}.
Finally, we note that this $\zeta(3)$ term in ABJM is not special to the round sphere free energy. Localization gives the $S^{3}$ free energy as a function of real masses and of the squashing parameter $b$~\cite{Nosaka:2015iiw, Chester:2021gdw, BenettiGenolini:2026kjy}. Its derivatives at $m=0$, $b=1$ fix integrated stress-tensor four-point functions, which have been used to determine the genus-zero $R^{4}$ vertex on $\mathrm{AdS}_{4}\times\mathbb{CP}^{3}$ with the same $\zeta(3)$ prefactor~\cite{Binder:2019mpb, Chester:2021gdw}. Since these deformations source bulk fields beyond the metric, they probe parts of the eight-derivative action that the round sphere does not and vice versa. Evaluating the IIA eight-derivative action on the corresponding backgrounds~\cite{Freedman:2013oja,Martelli:2011fu} would be a sharper test.

\section{Discussion}
\label{sec:discussion}
We derived the tree-level eight-derivative type IIA action for the metric, dilaton, and Ramond--Ramond one-form from the winding modes of one-loop 11d supergravity on a circle. In 10d flat space, the four-graviton term of \eqref{eq:iia_action} reduces to the tree-level $\zeta(3)R^{4}$ term~\cite{Gross:1986iv,Grisaru:1986vi}, and the full operator is consistent with maximal supersymmetry \cite{Herderschee:2026IIR4}, which has recently been shown to be much more constraining than previously appreciated~\cite{Elvang:2026pmc,Berman:2026ezk}. With four-form flux, the action reproduces the $\zeta(3)k^{2}$ term of the ABJM free energy obtained by localization. The same strategy should extend beyond $R^{4}$. For BFSS, the most useful next step is the next term in \eqref{eq:bfss_result}. Since the tree-level $\alpha'^{4}$ term vanishes, it is of order $T^{29/5}$ and comes from the tree-level $D^{4}R^{4}$ coupling at order $\alpha'^{5}$. Lattice fits for $a_1$ currently truncate at this order, so fixing it would remove a key systematic uncertainty. 

{\bf Note added}: after obtaining some of the initial results from AI, we learned that \cite{Compagnin:CompleteDbraneAction} obtained some higher derivative terms from M-theory in a similar way. As we were finishing our work, \cite{Kurlyand:2026cyz} appeared, which has some overlap with our ABJM result. \newline 

\section*{Acknowledgements} We thank Manki Kim, Silviu Pufu, Juan Maldacena, Douglas Stanford, and Xi Yin for helpful discussions. The work of A.H. was supported at MIT by the U.S. Department of Energy, Office of Science, Office of High Energy Physics under grant Contract Number DE-SC0012567. The work of Y.W. was supported by the NSF Grant 2609962 and by the Ambrose Monell Foundation and the Friends of the Institute for Advanced Study. 

\section*{AI statement}
The core proof strategy presented in this paper came largely from ChatGPT 6.1 Sol Ultra (OpenAI), and it performed all the computations leading to \eqref{eq:11d_compact}. However, many steps in the derivation were left unjustified, and these justifications were supplied by the authors.
We have included a PDF of the original output from 6.1 Sol in the arXiv source code for transparency.
\\
The authors also used Argon (Gemini), GPT-6 Astra (OpenAI) and Claude Opus (Anthropic) for formatting, editing and cross-checking. A.H.'s contribution was supported by complimentary access to Google Antigravity and frontier Gemini 4 models (Argon and Carbon) provided through Google's program to accelerate scientific workflows in Google Antigravity.

\appendix

\section{Evaluation of the BFSS free energy}
\label{app:BFSS}
In this appendix we give the details behind \eqref{eq:bfss_result}.

\subsection{Evaluation of bulk terms}
It is convenient to use the 11d uplift in unboosted coordinates \eqref{eq:d0_uplift_flat}. One can pick an orthonormal coframe
\begin{equation}
e^{0}=dt'~~,~~e^{1}=c\sqrt{h}dw~~,~~e^{2}=\frac{c}{\sqrt{h}}dr~~,~~e^{i}=c r\,\bar{e}^{i}
\end{equation}
with $\bar{e}^{i}$ an orthonormal coframe on the unit $S^{8}$. Up to the symmetries of the Riemann tensor, its non-zero frame components are
\begin{equation}
R_{12 12}=\frac{28r_{0}^{7}}{c^{2}r^{9}}, \qquad R_{ijkl}=\frac{r_{0}^{7}}{c^{2}r^{9}}(\delta_{ik}\delta_{jl}-\delta_{il}\delta_{jk}), \qquad R_{1i1i}=R_{2i2i}=-\frac{7 r_{0}^{7}}{2c^{2}r^{9}} \ .
\end{equation}

In evaluating $\hat{E}$ and $\hat{A}$ defined by \eqref{eq:iia_action} in this background, we obtain that they are of the form \eqref{eq:bfss_profiles}. To write the remaining constant matrices $E_{0}$ and $A_{0}$, we introduce the $SO(2)$ charge $\mathcal{Q}$, and the quadratic Casimirs of $SO(8)$ and $SO(10)$ as
\begin{equation}
\mathcal{Q}=-i J_{12}, \qquad \mathcal{C}_{8}=-\sum_{3\leq a<b\leq 10}J_{ab}^{2},\qquad\mathcal{C}_{10}=-\sum_{1\leq a<b\leq 10}J_{ab}^{2} \ .
\end{equation}
In terms of these commuting operators, we find that
\begin{equation}
E_{0}=\frac{63}{2} \mathcal{Q}^{2}+\frac{9}{2}\mathcal{C}_{8}-\frac{7}{2}\mathcal{C}_{10}~~,~~ A_{0}=\frac{7}{2}\mathcal{Q} \ .
\end{equation}

To compute the correction to the action, we need to evaluate the supertrace \eqref{eq:bfss_str}. We do so by first decomposing each field into irreducible representations of $SO(10)$, which fixes their eigenvalues under $\mathcal{C}_{10}$. In doing so, the supertrace can be rewritten in terms of 10d irreps as
\begin{equation}
\label{eq:Str10d}
\mathrm{Str}=\mathrm{tr}_{\mathbf{54}}+\mathrm{tr}_{\mathbf{120}}-\mathrm{tr}_{\mathbf{\mathbf{45}}}-\mathrm{tr}_{\mathbf{1}}-\text{tr}_{\mathbf{144}_{+}}+\text{tr}_{\mathbf{16}_{+}}
\end{equation}
with the bolded numbers $SO(10)$ irreps, and where we identified the two different chiralities, since our observable is not sensitive to them. We then further decompose each of these representations under $SO(2) \times SO(8)$, so that $\mathcal{C}_{8}$ and $\mathcal{Q}$ are fixed by the $SO(8)$ representation and the $SO(2)$ charge. 

Therefore, the left-hand side of \eqref{eq:bfss_str} reduces to a sum over Casimir eigenvalues, weighted by their appropriate multiplicity in $SO(10)$. It can be evaluated following the decomposition in \eqref{eq:Str10d} as
\begin{equation}
\begin{gathered}
\label{qval}
Q=\mathrm{Str}[(E_{0}-A_{0}^{2})^{3}(E_{0}-9A_{0}^{2})]= \\\underbrace{\frac{594057177}{16}}_{\mathbf{54}}
 + \underbrace{\frac{798919359}{32}}_{\mathbf{120}}
 - \underbrace{\frac{161183897}{16}}_{\mathbf{45}}
+ \underbrace{\frac{51883209}{4096}}_{\mathbf{16}_{+}}
 - \underbrace{\frac{226198100401}{4096}}_{\mathbf{144}_{+}}= -\frac{1633622445}{512}
\end{gathered}
\end{equation}
In particular, \eqref{eq:Str10d} shows that the final answer comes from a somewhat large cancellation between the bosonic and fermionic pieces, since the final result is about 17 times smaller in magnitude than either term. By using the value of $Q$, one can compute the leading correction to the free energy by evaluating the action \eqref{eq:iia_action} in the unperturbed BFSS background. This is valid because the first-order shift in the background does not modify the action at leading order, since we are expanding around a solution of the unperturbed equation of motion.

\subsection{Boundary terms}
One might be worried about the effect of boundary terms. At fixed inverse temperature $\beta$ and D0 charge $N$, we only need to evaluate the new eight-derivative terms on the leading D0 solution. 
Let us package all the bulk fields together as $\psi$. Then, including backreaction due to the new 8-derivative terms, $\psi =\psi_0+\varepsilon\psi_1+O(\varepsilon^2)$, with $\varepsilon\propto\alpha'^3$. Let $B_0$ and $B_8$ denote the old and new boundary terms, respectively. The full Euclidean action is
\begin{align}
 I_E^{\rm can}[\psi_\varepsilon]
 &= I_0^{\rm can}[\psi_0]
  + \varepsilon\Bigl[
      I_{8,\rm bulk}[\psi_0]+B_8[\psi_0]
      +\underbrace{\delta_{\psi_1}(I_{0,\rm bulk}+B_0)[\psi_0]}_{0}
    \Bigr]+O(\varepsilon^2).
\end{align}
The leading equations of motion make the interior variation of $I_{0,\rm bulk}$ vanish. Its surface variation cancels against $\delta_{\psi_1}B_0$ because $\psi_1$ preserves the prescribed boundary data. 

$B_0$ is chosen so that its variation cancels the outer surface term in
$\delta I_{0,\rm bulk}$ for any variation that preserves the prescribed
boundary data, as $\psi_1$ does. The same requirement applies to $B_8$:
to find its variation, we compute the outer surface term of the bulk action $\delta I_{8,\rm bulk} =  - \delta B_8$. 
Let us denote the bulk fields as $\psi^i=(A,B,\phi)$, where $g_{\tau\tau}=e^{2A}$,
$g_{rr}=e^{2B}$, and $\phi$ is the dilaton. Here we have eliminated the gauge fields using Gauss's law for fixed $N$. The interaction has eight
derivatives in total, but each field appears with at most two radial
derivatives. For static fields, 
\begin{equation}
I_{8,\rm bulk}=\beta\int_{r_0}^{R}dr\,L_8 \ .
\end{equation}
Varying the fields at fixed $\beta$ and integrating by parts in $r$ gives the
outer boundary contribution
\begin{align}
 \left.\delta I_{8,\rm bulk}\right|_{\rm outer}
 &= \beta\,\Theta_8(R) \nonumber\\
 &= \beta\left.\sum_i \left[
    \left(\frac{\partial L_8}{\partial \psi^{i\prime}}
      - \frac{d}{dr}\frac{\partial L_8}{\partial \psi^{i\prime\prime}}\right)
      \delta\psi^i
    + \frac{\partial L_8}{\partial \psi^{i\prime\prime}}\,
      \delta\psi^{i\prime}
    \right]\right|_{r=R},
\end{align}
where primes denote radial derivatives.

The boundary contribution to the free energy is $\varepsilon B_8/\beta$, so only the temperature dependence of $B_8/\beta$ matters for the entropy.
Since $T\propto r_0^{5/2}$ at fixed $N$, we can differentiate with respect to $r_0$. Because the boundary action is local, this measures the change in $B_8/\beta$ along the leading D0 family. The metric depends on $r_0$ through $1-(r_0/r)^7$, so $\delta A=-\delta B=O(R^{-7})$ and $\delta A'=O(R^{-8})$, while the dilaton is independent of $r_0$. For the interaction used here, the coefficients of $\delta A$ and $\delta B$ are $O(R^{-5/2})$, and the coefficient of $\delta A'$ is $O(R^{-3/2})$. Both contributions thus have the same leading power at fixed $r_0$:
\begin{align}
 \Theta_8(R) &\propto R^{-19/2} \qquad (R\gg r_0).
\end{align}
The variation of $B_8/\beta$ must cancel $\Theta_8(R)$, so $\partial_{\log r_0}(B_8/\beta)$ vanishes as $R\to\infty$.

The D0 geometry is not reliable when the radius $R$ is large because the string-frame curvature blows up. If we naively cut off the geometry at some radius $R_s$ where the curvature is of order the string scale, this surface variation is smaller than the corresponding variation of the eight-derivative bulk term by a factor of order $(r_0/R_s)^{19/2}\sim(T/\lambda^{1/3})^{19/5}$. At low temperatures, such a term is sub-leading to the bulk $\alpha'^3$ correction to the free energy.

\section{One-loop determinant on \texorpdfstring{$\mathrm{AdS}_{4}\times S^{7}/\mathbb{Z}_{k}$}{AdS4 x S7/Zk}}
\label{app:abjm}

In this appendix we derive \eqref{eq:abjm_result} from the one-loop determinant of 11d supergravity on the background \eqref{eq:abjm_uplift}. The strategy is the same as for the toy model and for the Ricci-flat backgrounds of the main text: we integrate out the KK tower on a small circle, trade the KK sum for a winding sum, and keep the local term with eight derivatives. Most of this appendix is devoted to the two new ingredients, the four-form flux and the curvature of the base.

\subsection*{The small circle} The Hopf fiber of $S^{7}/\mathbb{Z}_{k}$ plays the role of the thermal or small compactification circle with periodic spin structure for the fermions. Writing \eqref{eq:abjm_uplift} in the form \eqref{eq:11d_metric} with $z=\psi$, the fiber has constant radius $\rho=R/k$ and size $\ell=2\pi R/k$. The KK photon is $C=k\omega$, and its field strength is proportional to the K\"ahler form of $\mathbb{CP}^{3}$. Since $\rho$ is constant, only the field-strength term of \eqref{eq:toy_A_KK} survives, and the spin connection matrix $A$, which encodes the Lorentz rotation picked up around the fiber and should not be confused with $C$, is constant,
\begin{equation}
\label{eq:abjm_A}
A=\frac{1}{R}\big(J_{45}+J_{67}+J_{89}\big),
\end{equation}
where $4,\dots,9$ are orthonormal frame directions on $\mathbb{CP}^{3}$, paired by the complex structure. In the IIA regime, the circle is much smaller than the rest of the geometry, $\ell/R=2\pi/k\ll1$, so the expansion in $\ell$ of the effective action is well defined.

Every eight-derivative term is of order $R^{-8}\ell^{-3}$. On the Ricci-flat backgrounds of the main text, which suffice for the D0-brane application, only the terms built from $E$ and $A$ contribute at this order. Every other term, such as the curvature of the base or a derivative of $A$, carries fewer Lorentz generators than derivatives and vanishes inside $\mathrm{Str}$. With flux this counting fails: the flux enters the operators through the first-order terms $C^{a}$ below and through $E$, and these are not products of Lorentz generators. We therefore need to keep all eight-derivative terms, including those from the curvature of $H^{4}$ and $S^{7}$.

As in the main text, terms with fewer derivatives cancel by supersymmetry, and the gapless $n=0$ modes, here type IIA supergravity on $\mathrm{AdS}_{4}\times\mathbb{CP}^{3}$, do not contribute to the local term. They give the $\log k$ terms and zero-mode constants of the full one-loop answer (see e.g.~\cite{Kurlyand:2026cyz}).

\subsection*{Kinetic operators} It is convenient to work in Euclidean signature,
\begin{equation}
\label{eq:abjm_euclid}
ds_{11}^{2}=\frac{R^{2}}{4}ds_{H^{4}}^{2}+R^{2}ds_{S^{7}/\mathbb{Z}_{k}}^{2},\qquad G_{4}=\frac{3i R^3}{8}\,\mathrm{vol}_{H^{4}},
\end{equation}
where $ds_{H^{4}}^{2}$ and $ds_{S^{7}/\mathbb{Z}_{k}}^{2}$ have unit radius and $\mathrm{vol}_{H^{4}}$ is the corresponding normalized volume form. The flux is imaginary because it is electric (one of its legs is along Euclidean time). We use the Euclidean action
\begin{equation}
\label{eq:abjm_action}
S_{E}=\frac{1}{2\kappa_{11}^{2}}\Big[-\int R^{(11)}\ast1+\frac12\int G_{4}\wedge\ast G_{4}+\frac{i}{6}\int C_{3}\wedge G_{4}\wedge G_{4}\Big]\,,
\end{equation}
and on this background \eqref{eq:abjm_euclid} we have $G_{4}^{2}=-864/R^{2}$ and $R^{(11)}=-6/R^{2}$.  

We fix the de Donder gauge $\nabla^{N}h_{MN}-\frac12\nabla_{M}h^{N}{}_{N}=0$ for the metric fluctuation~\cite{tHooft:1974toh,Christensen:1979iy}, the Lorenz gauge $d^{\dagger}c_{3}=0$ for the three-form fluctuation, with its tower of ghosts for ghosts~\cite{Siegel:1980jj}, and the gamma-trace gauge $\Gamma^{M}\psi_{M}=\zeta$ for the gravitino, averaged  over $\zeta$ with a weight containing the Dirac operator $D=\Gamma^{M}\nabla_{M}$~\cite{Nielsen:1978mp,Kallosh:1978de}. These gauges make every kinetic operator of Laplace type, $-\nabla^{2}$ plus terms with fewer derivatives, as required by the heat kernel calculation below, and are standard in one-loop computations in supergravity on $\mathrm{AdS}_{4}\times X_{7}$~\cite{Bhattacharyya:2012ye} (see also~\cite{Sen:2012kpz}). Note that the ghost sectors do not mix: although a diffeomorphism shifts $c_{3}$ by $\iota_{\xi}G_{4}$, the de Donder condition is invariant under the three-form gauge symmetry, so the Faddeev--Popov matrix is triangular.

The one-loop effective action is then $\Gamma=\frac12\sum_{s}w_{s}\,\mathrm{Tr}'\log\Delta_{s}$, where the prime removes zero modes and the sectors are listed in Table~\ref{tab:abjm_sectors}.
\begin{table}[t]\centering
\begin{tabular}{lccr}
Sector & Rank & $w_{s}$ & Contribution to $j$\\ \hline
Metric and three-form & 231 & $1$ & $236177152/175$\\
Diffeomorphism ghost & 11 & $-2$ & $41996288/525$\\
Two-form ghost & 55 & $-2$ & $-125994496/105$\\
One-form ghost & 11 & $3$ & $61657856/175$\\
Scalar ghost & 1 & $-4$ & $-212992/75$\\
Gravitino & 352 & $-\tfrac12$ & $-8149658021/8400$\\
Supersymmetry ghost & 32 & $1$ & $-29237248/525$\\
Nielsen--Kallosh ghost & 32 & $\tfrac12$ & $36956581/8400$\\ \hline
Total & & & $-442368$
\end{tabular}
\caption{Sectors of the gauge-fixed one-loop determinant $\Gamma$ on $\mathrm{AdS}_{4}\times S^{7}/\mathbb{Z}_{k}$: rank of the fluctuating field, weight $w_{s}$ in $\Gamma$, and contribution to the dimensionless eight-derivative density $j=R^{8}I_{8}$ of \eqref{eq:abjm_gamma8}.}
\label{tab:abjm_sectors}
\end{table}
The ghost weights $-2$, $3$, and $-4$ arise from the tower of ghosts and ghosts-for-ghosts required to fix the reducible gauge symmetry of the three-form~\cite{Siegel:1980jj}. The fermion weights carry a factor $\frac12$ from the Majorana condition: a Majorana field contributes $(\det D)^{1/2}$ where $D=\Gamma^{M}\nabla_{M}$ is the Dirac operator, that is $-\frac14\mathrm{Tr}\log\Delta$ with $\Delta=-D^{2}$. The supersymmetry ghosts are two commuting Majorana spinors.
A third commuting Majorana spinor, the Nielsen--Kallosh ghost~\cite{Nielsen:1978mp,Kallosh:1978de}, compensates for the background dependence of the average over $\zeta$. Since all three ghosts are commuting spinors, their weights have the opposite sign to that of the gravitino.
The latter arises because we average over the gravitino gauge condition with a weight containing the Dirac operator, and must compensate for the background dependence of that average. The ghosts therefore enter with the opposite sign. The weights reduce to those of \eqref{eq:11d_func_str} in the absence of flux.

The flux enters the kinetic operators $\Delta_s$ in three ways. First, it mixes the metric and three-form fluctuations, $h_{MN}$ and $c_{3}$. Expanding $G_{4}\wedge\ast G_{4}$ to second order gives terms such as 
\begin{equation}
G_{4}\wedge\ast G_{4}\ni h^{MN}G_{M}^{PQR}(dc_{3})_{NPQR} \ ,
\end{equation}
with one derivative, so these two fields must be treated as a single sector of rank $66+165=231$. Second, it couples to the gravitino through 
\begin{equation}
\bar\psi_{M}\big(\Gamma^{MNPQRS}+12\,g^{MP}\Gamma^{QR}g^{NS}\big)\psi_{N}\,G_{PQRS}\,.
\end{equation}
Third, the Chern--Simons term in \eqref{eq:abjm_action} gives the quadratic term $\frac{i}{2\kappa_{11}^{2}}\int c_{3}\wedge G_{4}\wedge dc_{3}$, up to a total derivative, which is first order in derivatives and acts within the three-form block.
After gauge fixing, the gravitino and its ghosts have first-order operators $D+V$, where $D$ is the Dirac operator above, built from the Levi-Civita connection $\nabla$, and $V$ is linear in $G_{4}$. In the sign convention of the Chern--Simons term in  \eqref{eq:abjm_action}, supersymmetry selects the representation of the 11d Clifford algebra with $\Gamma^{0}\cdots\Gamma^{10}=-i$. With this choice $D+V$ is anti-Hermitian, so the second-order operator $-(D+V)^{2}$ is positive.
In all cases, the second-order kinetic operators take the form
\begin{equation}
\Delta=-(\nabla+C)^{2}+E,
\end{equation}
where the constant matrices $C^{a}$ collect the flux terms (first-order in derivatives) and act like an additional connection, and $E$ is a constant matrix built from the Riemann tensor and the flux. The matrix $A$ is not part of $C$: the spin connection is contained in $\nabla$. It enters only through the twist $e^{-w\ell A}$ below, because the KK momentum is defined by the Lie derivative along the Hopf fiber, which differs from $\nabla$ along the fiber by $\rho A$, as in \eqref{eq:toy_nabla}. Without flux, $C^{a}=0$ and $E$ reduces to \eqref{eq:11d_delta}.

\subsection*{Heat kernel} As in \eqref{eq:11d_winding_sum}, we write 
\begin{equation}
\log\Delta=-\int_{0}^{\infty}\frac{dt}{t}e^{-t\Delta}
\end{equation}
After Poisson resummation, the winding-$w$ term involves the heat kernel 
\begin{equation}
K(t;x,x')=\langle x|e^{-t\Delta}|x'\rangle
\end{equation}
between a point and its image a distance $|w|\ell$ along the Hopf fiber, rotated by $e^{-w\ell A}$ due to the spin connection $A$. For a Ricci-flat background with constant $E$ and $A$, this kernel is a Gaussian times $e^{-tE}$, which gives \eqref{eq:11d_winding_sum}. Here, the base is curved and $C^{a}$ does not commute with $E$, so we use the short-time expansion of the heat kernel at finite separation~\cite{Vassilevich:2003xt}.  This suffices for the local term: each power $t^{n}$ of the kernel gives a definite power of $w\ell$, and the eight-derivative term is the coefficient of $(|w|\ell)^{-3}$, as shown below.

Let $r$ and $s$ be the geodesic distances from the evaluation point in $H^{4}$ and $S^{7}$, with $s$ along the Hopf fiber. Then
\begin{equation}\label{Ktova}
K(t;r,s)=\frac{e^{-(r^{2}+s^{2})/(4t)}}{(4\pi t)^{11/2}}\,v(r,s)\sum_{n\geq0}t^{n}a_{n}(r,s),
\end{equation}
where 
\begin{equation}
v=(r/s_{\kappa_{A}}(r))^{3/2}(s/s_{\kappa_{S}}(s))^{3}
\end{equation}
is the square root of the Van Vleck determinant of the two constant-curvature factors~\cite{Visser:1992pz,Camporesi:1990wm,Vassilevich:2003xt}, with $s_{\kappa}(x)=\sin(\sqrt{\kappa}x)/\sqrt{\kappa}$ and sectional curvatures $\kappa_{A}=-4/R^{2}$, $\kappa_{S}=1/R^{2}$.  The exponents of $r$ and $s$ in $v$ follow from the general formula $(d-1)/2$ applied for the dimensions $d=4$ and $7$ relevant here. The matrix coefficients $a_{n}$ obey the following recursion equations, which follow from inserting this ansatz \eqref{Ktova} into the heat equation $(\partial_{t}+\Delta)K=0$ and matching powers of $t$,
\begin{equation}
\label{eq:abjm_transport}
(n+r\partial_{r}+s\partial_{s}+rC_{r}+sC_{s})\,a_{n}=v^{-1}\big((\nabla+C)^{2}-E\big)\big(v\,a_{n-1}\big),
\end{equation}
with 
\begin{equation}
a_{0}=e^{-rC_{r}-sC_{s}} \ ,
\end{equation}
which is the parallel transport with the extra connection $C$ (i.e. $C_{r}$ and $C_{s}$ are the components of $C$ along the two geodesics). 
In the limit where $C$ and $A$ commute, the trace in \eqref{eq:abjm_Tn} below therefore contains $e^{s(A-C_{s})}$, the transport of $-(\partial_{s}-A+C_{s})^{2}$ in the invariant frame.
Expanding $a_{n}$ in powers of $r$ and $s$ turns these equations into a finite recursion for constant matrices. Finally, we set $r=0$ and $s=|w|\ell$. The dependence on $r$ is only needed to compute the derivatives on the right-hand side.

\subsection*{Winding sum} Each term $t^{n}$ of the kernel, at separation $s=|w|\ell$, gives the proper-time integral
\begin{equation}
\label{eq:abjm_tint}
\int_{0}^{\infty}\frac{dt}{t}\,\frac{t^{n}\,e^{-(w\ell)^{2}/(4t)}}{(4\pi t)^{11/2}}=\frac{\Gamma(\frac{11}{2}-n)}{(4\pi)^{11/2}}\Big(\frac{4}{(w\ell)^{2}}\Big)^{\frac{11}{2}-n},
\end{equation}
as in the footnote to \eqref{eq:11d_compact}. A term $t^{n}s^{m}$ therefore comes with $(|w|\ell)^{2n+m-11}$, and the eight-derivative term, which carries $(|w|\ell)^{-3}$, has $m=8-2n$. It collects the coefficient of $s^{8-2n}$ in
\begin{equation}
\label{eq:abjm_Tn}
T_{n}(s)=\sum_{\sigma}w_{\sigma}\,\mathrm{tr}_{\sigma}\big[e^{sA}\,v(0,s)\,a_{n}(0,s)\big],
\end{equation}
where $\sigma$ runs over the sectors in Table~\ref{tab:abjm_sectors}. The sum over $w\neq0$ of the factor $(|w|\ell)^{-3}$ from \eqref{eq:abjm_tint} then gives $2\zeta(3)/\ell^{3}$, and the result has the normalization of \eqref{eq:11d_compact},
\begin{equation}
\label{eq:abjm_gamma8}
\Gamma_{8}=-\frac{\zeta(3)}{12288\,\pi^{5}}\int d\mathrm{vol}_{11}\,\ell^{-3}I_{8},\quad I_{8}=\sum_{n=0}^{4}W_{n}  T_{n}(s)|_{s^{8-2n}},
\end{equation}
where 
\begin{equation}
W_{n}=12288\,\Gamma(\frac{11}{2}-n)/(4^{n}\sqrt{\pi})=(362880,20160,1440,144,24)
\end{equation}
and $T_n(s)|_{s^m}$ extracts the coefficient of $s^{m}$ in $T_n(s)$. For a Ricci-flat background without flux, $C^{a}=0$, the curvature of the base drops out by the generator counting below \eqref{eq:11d_func_str}, and the part of $a_{n}$ that survives the supertrace at eight derivatives is
\begin{equation}
a_{n}=(-E)^{n}/n! \ .
\end{equation}
The same counting sets $v$ to $1$, since the curvature corrections in $v$ carry no Lorentz generators. Then $I_{8}$ recovers \eqref{eq:11d_compact}.

\subsection*{Evaluation of the density} The dimensionless density 
\begin{equation}
j=R^{8}I_{8}
\end{equation}
receives the contributions listed in the last column of Table~\ref{tab:abjm_sectors}, which sum to $j=-442368$. The input matrices $C^{a}$ and $E$ were computed numerically to high precision and rationalized. The recursion \eqref{eq:abjm_transport} was then run in exact arithmetic.\footnote{Working with $R=1$, all inputs, including the rotation generators and the Taylor coefficients of $v$, have entries in $\mathbb{Q}(i)$. We ran the recursion modulo several large primes and reconstructed the rational result by the Chinese remainder theorem. A floating-point run of the same recursion agrees with $j=-442368$ to $10^{-12}$ relative accuracy. No coefficient was fitted to the expected answer. Both the numerical and exact evaluations were assisted by Claude Opus 5.5.}
 
The analogous terms with zero, two, four and six derivatives cancel in the sum, as required by supersymmetry. The individual sectors depend on the gauge fixing and only the sum is meaningful. The density is constant, so the integral in \eqref{eq:abjm_gamma8} only needs the volumes and the circle length,
\begin{equation}
\mathrm{Vol}_{\rm ren}(H^{4}_{R/2})=\frac{\pi^{2}R^{4}}{12},\qquad \mathrm{Vol}(S^{7}_{R}/\mathbb{Z}_{k})=\frac{\pi^{4}R^{7}}{3k},\qquad \ell^{-3}=\frac{k^{3}}{8\pi^{3}R^{3}}.
\end{equation}
The powers of $R$ cancel, and 
\begin{equation}
\int d\mathrm{vol}_{11}\,\ell^{-3}I_{8}=\pi^{3}k^{2}j/288 \ .
\end{equation}
Evaluating $\Gamma_8$ in \eqref{eq:abjm_gamma8} gives the contribution to the free energy
\begin{equation}
\delta F_{\rm bulk}=-\frac{\zeta(3)\,k^{2}\,j}{3538944\,\pi^{2}}=\frac{\zeta(3)}{8\pi^{2}}\,k^{2}.
\end{equation}

\subsection*{Spectral check} The coefficient in \eqref{eq:abjm_result} also follows from the KK spectrum on $\mathrm{AdS}_{4}$. On $\mathrm{AdS}_{4}$, the continuous spectrum, used for the full KK tower in Ref.~\cite{Liu:2016dau}, misses some normalizable modes. 
Throughout, fluctuations and gauge parameters are required to be square-integrable on $H^{4}$, and the free energy is computed at fixed $N$.
The three-form modes $c_{3}=\sum_{I}\eta_{I}\wedge a_{I}$, with $\eta_{I}$ the normalizable harmonic two-forms on $H^{4}$ and $a_{I}$ one-forms on $S^{7}/\mathbb{Z}_{k}$, are not pure gauge, because the primitive of $\eta_{I}$ is not normalizable. Each harmonic mode gives $\frac12\log\det\Delta_{1,\perp}-\frac12\log\det{}'\Delta_{0}$ on $S^{7}/\mathbb{Z}_{k}$, where $\perp$ denotes coexact one-forms and the second term is the gauge symmetry $\delta a_{I}=d\lambda_{I}$ left on $S^{7}/\mathbb{Z}_{k}$. The number of harmonic modes is infinite, but after regularization with the renormalized volume of $H^{4}$ it is one~\cite{Bhattacharyya:2012ye}. The terms that grow with $k$ are listed in Table~\ref{tab:abjm_spectral}. The remaining terms, $\log k$, a constant, and terms that decay with $k$, are not part of the local eight-derivative term: the first two come from the gapless modes and zero modes, and the decaying ones from terms with more derivatives, which are of higher genus. They are discussed in Ref.~\cite{Kurlyand:2026cyz}. 
\begin{table}[t]\centering
\begin{tabular}{lccc}
 & $\zeta(7)k^{6}/\pi^{6}$ & $\zeta(5)k^{4}/\pi^{4}$ & $\zeta(3)k^{2}/\pi^{2}$\\ \hline
Continuous~\cite{Liu:2016dau} & $25/1024$ & $-3/128$ & $-1/18$\\
Harmonic & $-25/1024$ & $3/128$ & $13/72$\\ \hline
Sum & $0$ & $0$ & $1/8$
\end{tabular}
\caption{Terms with positive powers of $k$ in the one-loop free energy on $\mathrm{AdS}_{4}\times S^{7}/\mathbb{Z}_{k}$, from the continuous spectrum on $\mathrm{AdS}_{4}$ and from the normalizable harmonic two-form modes.}
\label{tab:abjm_spectral}
\end{table}
The harmonic modes cancel the $k^{6}$ and $k^{4}$ terms, which have no counterpart in the dual field theory, and the sum reproduces \eqref{eq:abjm_result}. Unlike the spectral computation, the heat-kernel computation above needs no separate harmonic-mode contribution: \eqref{Ktova} is the kernel of $e^{-t\Delta}$ on all normalizable fluctuations, harmonic modes included, so $\delta F_{\rm bulk}$ corresponds to the sum in Table~\ref{tab:abjm_spectral}.

\bibliographystyle{JHEP}
\bibliography{references}

@article{Maldacena:1997re,
    author = "Maldacena, Juan Martin",
    title = "{The Large $N$ limit of superconformal field theories and supergravity}",
    eprint = "hep-th/9711200",
    archivePrefix = "arXiv",
    reportNumber = "HUTP-97-A097, HUTP-98-A097",
    doi = "10.4310/ATMP.1998.v2.n2.a1",
    journal = "Adv. Theor. Math. Phys.",
    volume = "2",
    pages = "231--252",
    year = "1998"
}

@unpublished{Compagnin:CompleteDbraneAction,
  author = {Compagnin, Federico and Kim, Manki and Schachner, Andreas and Schreyer, Simon},
  title  = {The complete {D-brane} action at order {$(\alpha')^2$}},
  note   = {Unpublished manuscript}
}

@article{Gubser:1998bc,
    author = "Gubser, S. S. and Klebanov, Igor R. and Polyakov, Alexander M.",
    title = "{Gauge theory correlators from noncritical string theory}",
    eprint = "hep-th/9802109",
    archivePrefix = "arXiv",
    reportNumber = "PUPT-1767",
    doi = "10.1016/S0370-2693(98)00377-3",
    journal = "Phys. Lett. B",
    volume = "428",
    pages = "105--114",
    year = "1998"
}

@article{Witten:1998qj,
    author = "Witten, Edward",
    title = "{Anti de Sitter space and holography}",
    eprint = "hep-th/9802150",
    archivePrefix = "arXiv",
    reportNumber = "IASSNS-HEP-98-15",
    doi = "10.4310/ATMP.1998.v2.n2.a2",
    journal = "Adv. Theor. Math. Phys.",
    volume = "2",
    pages = "253--291",
    year = "1998"
}

@article{Banks:1996vh,
    author = "Banks, Tom and Fischler, W. and Shenker, S. H. and Susskind, Leonard",
    title = "{M theory as a matrix model: A conjecture}",
    eprint = "hep-th/9610043",
    archivePrefix = "arXiv",
    reportNumber = "RU-96-95, SU-ITP-96-12, UTTG-13-96",
    doi = "10.1103/PhysRevD.55.5112",
    journal = "Phys. Rev. D",
    volume = "55",
    pages = "5112--5128",
    year = "1997"
}

@article{Itzhaki:1998dd,
    author = "Itzhaki, Nissan and Maldacena, Juan Martin and Sonnenschein, Jacob and Yankielowicz, Shimon",
    title = "{Supergravity and the large N limit of theories with sixteen supercharges}",
    eprint = "hep-th/9802042",
    archivePrefix = "arXiv",
    reportNumber = "TAUP-2474-98, HUTP-98-A003",
    doi = "10.1103/PhysRevD.58.046004",
    journal = "Phys. Rev. D",
    volume = "58",
    pages = "046004",
    year = "1998"
}

@article{Berkowitz:2016jlq,
    author = "Berkowitz, Evan and Rinaldi, Enrico and Hanada, Masanori and Ishiki, Goro and Shimasaki, Shinji and Vranas, Pavlos",
    title = "{Precision lattice test of the gauge/gravity duality at large-$N$}",
    eprint = "1606.04951",
    archivePrefix = "arXiv",
    primaryClass = "hep-lat",
    reportNumber = "LLNL-JRNL-694385, UTHEP-689, YITP-16-73",
    doi = "10.1103/PhysRevD.94.094501",
    journal = "Phys. Rev. D",
    volume = "94",
    number = "9",
    pages = "094501",
    year = "2016"
}

@article{Pateloudis:2022ijr,
    author = {Pateloudis, Stratos and Bergner, Georg and Hanada, Masanori and Rinaldi, Enrico and Sch{\"a}fer, Andreas and Vranas, Pavlos and Watanabe, Hiromasa and Bodendorfer, Norbert},
    collaboration = "Monte Carlo String/M-theory (MCSMC)",
    title = "{Precision test of gauge/gravity duality in D0-brane matrix model at low temperature}",
    eprint = "2210.04881",
    archivePrefix = "arXiv",
    primaryClass = "hep-th",
    doi = "10.1007/JHEP03(2023)071",
    journal = "JHEP",
    volume = "03",
    pages = "071",
    year = "2023"
}

@article{Halimeh:2024bth,
    author = {Halimeh, Jad C. and Hanada, Masanori and Matsuura, Shunji and Nori, Franco and Rinaldi, Enrico and Sch{\"a}fer, Andreas},
    title = "{A universal framework for the quantum simulation of Yang--Mills theory}",
    eprint = "2411.13161",
    archivePrefix = "arXiv",
    primaryClass = "quant-ph",
    reportNumber = "RIKEN-iTHEMS-Report-24",
    doi = "10.1038/s42005-025-02421-6",
    journal = "Commun. Phys.",
    volume = "9",
    number = "1",
    pages = "67",
    year = "2026"
}

@article{Gharibyan:2020bab,
    author = "Gharibyan, Hrant and Hanada, Masanori and Honda, Masazumi and Liu, Junyu",
    title = "{Toward simulating superstring/M-theory on a quantum computer}",
    eprint = "2011.06573",
    archivePrefix = "arXiv",
    primaryClass = "hep-th",
    reportNumber = "CALT-TH-2020-013, DMUS-MP-20/10, YITP-20-146",
    doi = "10.1007/JHEP07(2021)140",
    journal = "JHEP",
    volume = "07",
    pages = "140",
    year = "2021"
}

@article{Wang:2015jna,
    author = "Wang, Yifan and Yin, Xi",
    title = "{Constraining Higher Derivative Supergravity with Scattering Amplitudes}",
    eprint = "1502.03810",
    archivePrefix = "arXiv",
    primaryClass = "hep-th",
    reportNumber = "MIT-CTP-4644",
    doi = "10.1103/PhysRevD.92.041701",
    journal = "Phys. Rev. D",
    volume = "92",
    number = "4",
    pages = "041701",
    year = "2015"
}

@article{Wang:2015aua,
    author = "Wang, Yifan and Yin, Xi",
    title = "{Supervertices and Non-renormalization Conditions in Maximal Supergravity Theories}",
    eprint = "1505.05861",
    archivePrefix = "arXiv",
    primaryClass = "hep-th",
    reportNumber = "MIT-CTP-4673",
    month = "5",
    year = "2015"
}

@article{Green:1999pu,
    author = "Green, Michael B. and Kwon, Hwang-h. and Vanhove, Pierre",
    title = "{Two loops in eleven-dimensions}",
    eprint = "hep-th/9910055",
    archivePrefix = "arXiv",
    reportNumber = "DAMTP-1999-120, CERN-TH-99-283, SACLAY-SPH-T-99-100",
    doi = "10.1103/PhysRevD.61.104010",
    journal = "Phys. Rev. D",
    volume = "61",
    pages = "104010",
    year = "2000"
}

@article{Rinaldi:2021jbg,
    author = "Rinaldi, Enrico and Han, Xizhi and Hassan, Mohammad and Feng, Yuan and Nori, Franco and McGuigan, Michael and Hanada, Masanori",
    title = "{Matrix-Model Simulations Using Quantum Computing, Deep Learning, and Lattice Monte Carlo}",
    eprint = "2108.02942",
    archivePrefix = "arXiv",
    primaryClass = "quant-ph",
    reportNumber = "RIKEN-iTHEMS-Report-21, DMUS-MP-21/10",
    doi = "10.1103/PRXQuantum.3.010324",
    journal = "PRX Quantum",
    volume = "3",
    number = "1",
    pages = "010324",
    year = "2022"
}

@article{Maldacena:2023acv,
    author = "Maldacena, Juan",
    title = "{A simple quantum system that describes a black hole}",
    eprint = "2303.11534",
    archivePrefix = "arXiv",
    primaryClass = "hep-th",
    month = "3",
    year = "2023"
}

@article{Cremmer:1978km,
    author = "Cremmer, E. and Julia, B. and Scherk, Joel",
    title = "{Supergravity Theory in 11 Dimensions}",
    reportNumber = "LPTENS-78-10",
    doi = "10.1016/0370-2693(78)90894-8",
    journal = "Phys. Lett. B",
    volume = "76",
    pages = "409--412",
    year = "1978"
}

@article{Witten:1995ex,
    author = "Witten, Edward",
    title = "{String theory dynamics in various dimensions}",
    eprint = "hep-th/9503124",
    archivePrefix = "arXiv",
    reportNumber = "IASSNS-HEP-95-18",
    doi = "10.1016/0550-3213(95)00158-O",
    journal = "Nucl. Phys. B",
    volume = "443",
    pages = "85--126",
    year = "1995"
}

@article{Townsend:1995kk,
    author = "Townsend, P. K.",
    title = "{The eleven-dimensional supermembrane revisited}",
    eprint = "hep-th/9501068",
    archivePrefix = "arXiv",
    reportNumber = "DAMTP-R-95-2",
    doi = "10.1016/0370-2693(95)00397-4",
    journal = "Phys. Lett. B",
    volume = "350",
    pages = "184--187",
    year = "1995"
}

@article{Green:1997as,
    author = "Green, Michael B. and Gutperle, Michael and Vanhove, Pierre",
    title = "{One loop in eleven dimensions}",
    eprint = "hep-th/9706175",
    archivePrefix = "arXiv",
    reportNumber = "DAMTP-97-58, CPTH-S-528-0697",
    doi = "10.1016/S0370-2693(97)00931-3",
    journal = "Phys. Lett. B",
    volume = "409",
    pages = "177--184",
    year = "1997"
}

@article{Green:1997di,
    author = "Green, Michael B. and Vanhove, Pierre",
    title = "{D instantons, strings and M theory}",
    eprint = "hep-th/9704145",
    archivePrefix = "arXiv",
    reportNumber = "DAMTP-97-31, CPTH-S501-0497",
    doi = "10.1016/S0370-2693(97)00785-5",
    journal = "Phys. Lett. B",
    volume = "408",
    pages = "122--134",
    year = "1997"
}

@article{Russo:1997mk,
    author = "Russo, J. G. and Tseytlin, Arkady A.",
    title = "{One loop four graviton amplitude in eleven-dimensional supergravity}",
    eprint = "hep-th/9707134",
    archivePrefix = "arXiv",
    reportNumber = "IMPERIAL-TP-96-97-53",
    doi = "10.1016/S0550-3213(97)00631-7",
    journal = "Nucl. Phys. B",
    volume = "508",
    pages = "245--259",
    year = "1997"
}

@article{Gross:1986iv,
    author = "Gross, David J. and Witten, Edward",
    title = "{Superstring Modifications of Einstein's Equations}",
    reportNumber = "Print-86-0250 (PRINCETON)",
    doi = "10.1016/0550-3213(86)90429-3",
    journal = "Nucl. Phys. B",
    volume = "277",
    pages = "1",
    year = "1986"
}

@article{Grisaru:1986vi,
    author = "Grisaru, Marcus T. and Zanon, D.",
    title = "{$\sigma$ Model Superstring Corrections to the Einstein-hilbert Action}",
    reportNumber = "HUTP-86/A046, BRX-TH-202",
    doi = "10.1016/0370-2693(86)90765-3",
    journal = "Phys. Lett. B",
    volume = "177",
    pages = "347--351",
    year = "1986"
}

@article{Green:1997tv,
    author = "Green, Michael B. and Gutperle, Michael",
    title = "{Effects of D instantons}",
    eprint = "hep-th/9701093",
    archivePrefix = "arXiv",
    reportNumber = "DAMTP-96-104",
    doi = "10.1016/S0550-3213(97)00269-1",
    journal = "Nucl. Phys. B",
    volume = "498",
    pages = "195--227",
    year = "1997"
}

@article{Green:1998by,
    author = "Green, Michael B. and Sethi, Savdeep",
    title = "{Supersymmetry constraints on type IIB supergravity}",
    eprint = "hep-th/9808061",
    archivePrefix = "arXiv",
    reportNumber = "DAMTP-98-96, IASSNS-HEP-98-70",
    doi = "10.1103/PhysRevD.59.046006",
    journal = "Phys. Rev. D",
    volume = "59",
    pages = "046006",
    year = "1999"
}

@article{Witten:1995im,
    author = "Witten, Edward",
    title = "{Bound states of strings and p-branes}",
    eprint = "hep-th/9510135",
    archivePrefix = "arXiv",
    reportNumber = "IASSNS-HEP-95-83",
    doi = "10.1016/0550-3213(95)00610-9",
    journal = "Nucl. Phys. B",
    volume = "460",
    pages = "335--350",
    year = "1996"
}

@article{Biggs:2023sqw,
    author = "Biggs, Anna and Maldacena, Juan",
    title = "{Scaling similarities and quasinormal modes of D0 black hole solutions}",
    eprint = "2303.09974",
    archivePrefix = "arXiv",
    primaryClass = "hep-th",
    doi = "10.1007/JHEP11(2023)155",
    journal = "JHEP",
    volume = "11",
    pages = "155",
    year = "2023"
}

@article{Klebanov:1996un,
    author = "Klebanov, Igor R. and Tseytlin, Arkady A.",
    title = "{Entropy of near extremal black p-branes}",
    eprint = "hep-th/9604089",
    archivePrefix = "arXiv",
    reportNumber = "PUPT-1613, IMPERIAL-TP-95-96-39",
    doi = "10.1016/0550-3213(96)00295-7",
    journal = "Nucl. Phys. B",
    volume = "475",
    pages = "164--178",
    year = "1996"
}

@article{Anagnostopoulos:2007fw,
    author = "Anagnostopoulos, Konstantinos N. and Hanada, Masanori and Nishimura, Jun and Takeuchi, Shingo",
    title = "{Monte Carlo studies of supersymmetric matrix quantum mechanics with sixteen supercharges at finite temperature}",
    eprint = "0707.4454",
    archivePrefix = "arXiv",
    primaryClass = "hep-th",
    reportNumber = "RIKEN-TH-112, KEK-TH-1165",
    doi = "10.1103/PhysRevLett.100.021601",
    journal = "Phys. Rev. Lett.",
    volume = "100",
    pages = "021601",
    year = "2008"
}

@article{Catterall:2008yz,
    author = "Catterall, Simon and Wiseman, Toby",
    title = "{Black hole thermodynamics from simulations of lattice Yang-Mills theory}",
    eprint = "0803.4273",
    archivePrefix = "arXiv",
    primaryClass = "hep-th",
    doi = "10.1103/PhysRevD.78.041502",
    journal = "Phys. Rev. D",
    volume = "78",
    pages = "041502",
    year = "2008"
}

@article{Hanada:2008ez,
    author = "Hanada, Masanori and Hyakutake, Yoshifumi and Nishimura, Jun and Takeuchi, Shingo",
    title = "{Higher derivative corrections to black hole thermodynamics from supersymmetric matrix quantum mechanics}",
    eprint = "0811.3102",
    archivePrefix = "arXiv",
    primaryClass = "hep-th",
    doi = "10.1103/PhysRevLett.102.191602",
    journal = "Phys. Rev. Lett.",
    volume = "102",
    pages = "191602",
    year = "2009"
}

@article{Kadoh:2015mka,
    author = "Kadoh, Daisuke and Kamata, Syo",
    title = "{Gauge/gravity duality and lattice simulations of one dimensional SYM with sixteen supercharges}",
    eprint = "1503.08499",
    archivePrefix = "arXiv",
    primaryClass = "hep-lat",
    month = "3",
    year = "2015"
}

@article{Filev:2015hia,
    author = "Filev, Veselin G. and O'Connor, Denjoe",
    title = "{The BFSS model on the lattice}",
    eprint = "1506.01366",
    archivePrefix = "arXiv",
    primaryClass = "hep-th",
    reportNumber = "DIAS-STP-15-09",
    doi = "10.1007/JHEP05(2016)167",
    journal = "JHEP",
    volume = "05",
    pages = "167",
    year = "2016"
}

@article{Hanada:2016zxj,
    author = "Hanada, Masanori and Hyakutake, Yoshifumi and Ishiki, Goro and Nishimura, Jun",
    title = "{Numerical tests of the gauge/gravity duality conjecture for D0-branes at finite temperature and finite N}",
    eprint = "1603.00538",
    archivePrefix = "arXiv",
    primaryClass = "hep-th",
    reportNumber = "YITP-16-20, UTHEP-682, KEK-TH-1891",
    doi = "10.1103/PhysRevD.94.086010",
    journal = "Phys. Rev. D",
    volume = "94",
    number = "8",
    pages = "086010",
    year = "2016"
}

@article{Hanada:2013rga,
    author = "Hanada, Masanori and Hyakutake, Yoshifumi and Ishiki, Goro and Nishimura, Jun",
    title = "{Holographic description of quantum black hole on a computer}",
    eprint = "1311.5607",
    archivePrefix = "arXiv",
    primaryClass = "hep-th",
    reportNumber = "YITP-13-96, KEK-TH-1686, YITP-13-96, KEK-TH-1686",
    doi = "10.1126/science.1250122",
    journal = "Science",
    volume = "344",
    pages = "882--885",
    year = "2014"
}

@article{Hyakutake:2013vwa,
    author = "Hyakutake, Yoshifumi",
    title = "{Quantum near-horizon geometry of a black 0-brane}",
    eprint = "1311.7526",
    archivePrefix = "arXiv",
    primaryClass = "hep-th",
    doi = "10.1093/ptep/ptu028",
    journal = "PTEP",
    volume = "2014",
    pages = "033B04",
    year = "2014"
}

@article{Nilsson:1984bj,
    author = "Nilsson, B. E. W. and Pope, C. N.",
    editor = "Salam, A. and Sezgin, E.",
    title = "{Hopf Fibration of Eleven-dimensional Supergravity}",
    reportNumber = "IMPERIAL-TP-83-84-41",
    doi = "10.1088/0264-9381/1/5/005",
    journal = "Class. Quant. Grav.",
    volume = "1",
    pages = "499",
    year = "1984"
}

@article{Aharony:2008ug,
    author = "Aharony, Ofer and Bergman, Oren and Jafferis, Daniel Louis and Maldacena, Juan",
    title = "{N=6 superconformal Chern-Simons-matter theories, M2-branes and their gravity duals}",
    eprint = "0806.1218",
    archivePrefix = "arXiv",
    primaryClass = "hep-th",
    reportNumber = "WIS-12-08-JUN-DPP",
    doi = "10.1088/1126-6708/2008/10/091",
    journal = "JHEP",
    volume = "10",
    pages = "091",
    year = "2008"
}

@article{Drukker:2010nc,
    author = "Drukker, Nadav and Marino, Marcos and Putrov, Pavel",
    title = "{From weak to strong coupling in ABJM theory}",
    eprint = "1007.3837",
    archivePrefix = "arXiv",
    primaryClass = "hep-th",
    reportNumber = "HU-EP-10-39",
    doi = "10.1007/s00220-011-1253-6",
    journal = "Commun. Math. Phys.",
    volume = "306",
    pages = "511--563",
    year = "2011"
}

@article{Herzog:2010hf,
    author = "Herzog, Christopher P. and Klebanov, Igor R. and Pufu, Silviu S. and Tesileanu, Tiberiu",
    title = "{Multi-Matrix Models and Tri-Sasaki Einstein Spaces}",
    eprint = "1011.5487",
    archivePrefix = "arXiv",
    primaryClass = "hep-th",
    reportNumber = "PUPT-2359",
    doi = "10.1103/PhysRevD.83.046001",
    journal = "Phys. Rev. D",
    volume = "83",
    pages = "046001",
    year = "2011"
}

@article{Emparan:1999pm,
    author = "Emparan, Roberto and Johnson, Clifford V. and Myers, Robert C.",
    title = "{Surface terms as counterterms in the AdS / CFT correspondence}",
    eprint = "hep-th/9903238",
    archivePrefix = "arXiv",
    reportNumber = "DTP-99-21, UK-99-04, MCGILL-99-12, EHU-FT-9906",
    doi = "10.1103/PhysRevD.60.104001",
    journal = "Phys. Rev. D",
    volume = "60",
    pages = "104001",
    year = "1999"
}

@article{Kapustin:2009kz,
    author = "Kapustin, Anton and Willett, Brian and Yaakov, Itamar",
    title = "{Exact Results for Wilson Loops in Superconformal Chern-Simons Theories with Matter}",
    eprint = "0909.4559",
    archivePrefix = "arXiv",
    primaryClass = "hep-th",
    reportNumber = "CALT-68-2750",
    doi = "10.1007/JHEP03(2010)089",
    journal = "JHEP",
    volume = "03",
    pages = "089",
    year = "2010"
}

@article{Fuji:2011km,
    author = "Fuji, Hiroyuki and Hirano, Shinji and Moriyama, Sanefumi",
    title = "{Summing Up All Genus Free Energy of ABJM Matrix Model}",
    eprint = "1106.4631",
    archivePrefix = "arXiv",
    primaryClass = "hep-th",
    doi = "10.1007/JHEP08(2011)001",
    journal = "JHEP",
    volume = "08",
    pages = "001",
    year = "2011"
}

@article{Marino:2011eh,
    author = "Marino, Marcos and Putrov, Pavel",
    title = "{ABJM theory as a Fermi gas}",
    eprint = "1110.4066",
    archivePrefix = "arXiv",
    primaryClass = "hep-th",
    doi = "10.1088/1742-5468/2012/03/P03001",
    journal = "J. Stat. Mech.",
    volume = "1203",
    pages = "P03001",
    year = "2012"
}

@article{Hanada:2012si,
    author = "Hanada, Masanori and Honda, Masazumi and Honma, Yoshinori and Nishimura, Jun and Shiba, Shotaro and Yoshida, Yutaka",
    title = "{Numerical studies of the ABJM theory for arbitrary N at arbitrary coupling constant}",
    eprint = "1202.5300",
    archivePrefix = "arXiv",
    primaryClass = "hep-th",
    reportNumber = "KEK-TH-1516, NSF-KITP-12-026, IMPERIAL-TP-2012-SS-01",
    doi = "10.1007/JHEP05(2012)121",
    journal = "JHEP",
    volume = "05",
    pages = "121",
    year = "2012"
}

@article{Bergman:2009zh,
    author = "Bergman, Oren and Hirano, Shinji",
    title = "{Anomalous radius shift in AdS(4)/CFT(3)}",
    eprint = "0902.1743",
    archivePrefix = "arXiv",
    primaryClass = "hep-th",
    doi = "10.1088/1126-6708/2009/07/016",
    journal = "JHEP",
    volume = "07",
    pages = "016",
    year = "2009"
}

@article{Aharony:2009fc,
    author = "Aharony, Ofer and Hashimoto, Akikazu and Hirano, Shinji and Ouyang, Peter",
    title = "{D-brane Charges in Gravitational Duals of 2+1 Dimensional Gauge Theories and Duality Cascades}",
    eprint = "0906.2390",
    archivePrefix = "arXiv",
    primaryClass = "hep-th",
    reportNumber = "MAD-TH-09-05, WIS-04-09-JUN-DPP",
    doi = "10.1007/JHEP01(2010)072",
    journal = "JHEP",
    volume = "01",
    pages = "072",
    year = "2010"
}

@article{Chester:2018aca,
    author = "Chester, Shai M. and Pufu, Silviu S. and Yin, Xi",
    title = "{The M-Theory S-Matrix From ABJM: Beyond 11D Supergravity}",
    eprint = "1804.00949",
    archivePrefix = "arXiv",
    primaryClass = "hep-th",
    reportNumber = "PUPT-2556",
    doi = "10.1007/JHEP08(2018)115",
    journal = "JHEP",
    volume = "08",
    pages = "115",
    year = "2018"
}

@article{Binder:2019mpb,
    author = "Binder, Damon J. and Chester, Shai M. and Pufu, Silviu S.",
    title = "{AdS$_{4}$/CFT$_{3}$ from weak to strong string coupling}",
    eprint = "1906.07195",
    archivePrefix = "arXiv",
    primaryClass = "hep-th",
    reportNumber = "PUPT-2591",
    doi = "10.1007/JHEP01(2020)034",
    journal = "JHEP",
    volume = "01",
    pages = "034",
    year = "2020"
}

@article{Elvang:2026pmc,
    author = "Elvang, Henriette and Herderschee, Aidan and Morales, Roger",
    title = "{String theory from maximal supersymmetry}",
    eprint = "2601.11705",
    archivePrefix = "arXiv",
    primaryClass = "hep-th",
    reportNumber = "LITP-25-18",
    doi = "10.1007/JHEP07(2026)105",
    journal = "JHEP",
    volume = "07",
    pages = "105",
    year = "2026"
}

@article{Berman:2026ezk,
    author = "Berman, Justin and Caron-Huot, Simon and Chandra, Aditi V. and Elvang, Henriette and Herderschee, Aidan and Lin, Loki L. and Morales, Roger",
    title = "{Gravitational Effective Theories with Maximal Supersymmetry and a Peculiar Parity}",
    eprint = "2607.14230",
    archivePrefix = "arXiv",
    primaryClass = "hep-th",
    reportNumber = "LITP-26-13",
    month = "7",
    year = "2026"
}

@article{Batra:2026tob,
    author = "Batra, Gauri and Lin, Henry W. and Tang, Haifeng",
    title = "{An effective field theory approach to the sign problem in BFSS}",
    eprint = "2606.18396",
    archivePrefix = "arXiv",
    primaryClass = "hep-th",
    month = "6",
    year = "2026"
}

@article{Beccaria:2023hhi,
    author = "Beccaria, Matteo and Tseytlin, Arkady A.",
    title = "{Comments on ABJM free energy on S3 at large N and perturbative expansions in M-theory and string theory}",
    eprint = "2306.02862",
    archivePrefix = "arXiv",
    primaryClass = "hep-th",
    reportNumber = "Imperial-TP-AT-2023-03",
    doi = "10.1016/j.nuclphysb.2023.116286",
    journal = "Nucl. Phys. B",
    volume = "994",
    pages = "116286",
    year = "2023"
}

@article{BenettiGenolini:2026kjy,
    author = "Benetti Genolini, Pietro and Gaar, Florian and Gauntlett, Jerome P. and Park, Jaeha and Sparks, James",
    title = "{Localizing the M-Theory Path Integral}",
    eprint = "2609.31829",
    archivePrefix = "arXiv",
    primaryClass = "hep-th",
    month = "sep",
    year = "2026"
}

@article{Berkovits:2004px,
    author = "Berkovits, Nathan",
    title = "{Multiloop amplitudes and vanishing theorems using the pure spinor formalism for the superstring}",
    eprint = "hep-th/0406055",
    archivePrefix = "arXiv",
    reportNumber = "IFT-P-020-2004",
    doi = "10.1088/1126-6708/2004/09/047",
    journal = "JHEP",
    volume = "09",
    pages = "047",
    year = "2004"
}

@article{Bobev:2026gir,
    author = "Bobev, Nikolay and Gautason, Fridrik Freyr and van Muiden, Jesse",
    title = "{Holographic Tests of the $\mu$ Ensemble}",
    eprint = "2607.06493",
    archivePrefix = "arXiv",
    primaryClass = "hep-th",
    month = "7",
    year = "2026"
}

@article{deWit:1988wri,
    author = "de Wit, B. and Hoppe, J. and Nicolai, H.",
    title = "{On the Quantum Mechanics of Supermembranes}",
    reportNumber = "THU-88-15, KA-THEP-6/88",
    doi = "10.1016/0550-3213(88)90116-2",
    journal = "Nucl. Phys. B",
    volume = "305",
    pages = "545",
    year = "1988"
}

@article{Fradkin:1982kf,
    author = "Fradkin, E. S. and Tseytlin, Arkady A.",
    title = "{Quantum Properties of Higher Dimensional and Dimensionally Reduced Supersymmetric Theories}",
    reportNumber = "LEBEDEV-82-265",
    doi = "10.1016/0550-3213(83)90022-6",
    journal = "Nucl. Phys. B",
    volume = "227",
    pages = "252",
    year = "1983"
}

@article{Garousi:2020gio,
    author = "Garousi, Mohammad R.",
    title = "{Effective action of type II superstring theories at order $\alpha'^{3}$: NS-NS couplings}",
    eprint = "2011.02753",
    archivePrefix = "arXiv",
    primaryClass = "hep-th",
    doi = "10.1007/JHEP02(2021)157",
    journal = "JHEP",
    volume = "02",
    pages = "157",
    year = "2021"
}

@article{Garousi:2025wfk,
    author = "Garousi, Mohammad R.",
    title = "{Type IIA supergravity at one loop: {\ensuremath{\alpha}}'$^{3}$ terms in the metric-dilaton-RR one-form sector}",
    eprint = "2508.10529",
    archivePrefix = "arXiv",
    primaryClass = "hep-th",
    doi = "10.1007/JHEP11(2025)105",
    journal = "JHEP",
    volume = "11",
    pages = "105",
    year = "2025"
}

@article{Gautason:2025plx,
    author = "Gautason, Fridrik Freyr and van Muiden, Jesse",
    title = "{Ensembles in M-theory and holography}",
    eprint = "2505.21633",
    archivePrefix = "arXiv",
    primaryClass = "hep-th",
    doi = "10.1007/JHEP11(2025)078",
    journal = "JHEP",
    volume = "11",
    pages = "078",
    year = "2025"
}

@article{Gubser:1998nz,
    author = "Gubser, Steven S. and Klebanov, Igor R. and Tseytlin, Arkady A.",
    title = "{Coupling constant dependence in the thermodynamics of N=4 supersymmetric Yang-Mills theory}",
    eprint = "hep-th/9805156",
    archivePrefix = "arXiv",
    reportNumber = "IASSNS-HEP-98-47, PUPT-1794, IMPERIAL-TP-97-98-47",
    doi = "10.1016/S0550-3213(98)00514-8",
    journal = "Nucl. Phys. B",
    volume = "534",
    pages = "202--222",
    year = "1998"
}

@article{Paulos:2008tn,
    author = "Paulos, Miguel F.",
    title = "{Higher derivative terms including the Ramond-Ramond five-form}",
    eprint = "0804.0763",
    archivePrefix = "arXiv",
    primaryClass = "hep-th",
    doi = "10.1088/1126-6708/2008/10/047",
    journal = "JHEP",
    volume = "10",
    pages = "047",
    year = "2008"
}

@article{Liu:2022bfg,
    author = "Liu, James T. and Minasian, Ruben and Savelli, Raffaele and Schachner, Andreas",
    title = "{Type IIB at eight derivatives: insights from Superstrings, Superfields and Superparticles}",
    eprint = "2205.11530",
    archivePrefix = "arXiv",
    primaryClass = "hep-th",
    doi = "10.1007/JHEP08(2022)267",
    journal = "JHEP",
    volume = "08",
    pages = "267",
    year = "2022"
}

@article{Liu:2025uqu,
    author = "Liu, James T. and Minasian, Ruben and Savelli, Raffaele and Schachner, Andreas",
    title = "{Type IIB at eight derivatives: five-point axio-dilaton couplings}",
    eprint = "2507.07934",
    archivePrefix = "arXiv",
    primaryClass = "hep-th",
    reportNumber = "LITP-25-10, LMU-ASC 19/25",
    doi = "10.1007/JHEP09(2025)158",
    journal = "JHEP",
    volume = "09",
    pages = "158",
    year = "2025"
}

@article{Nosaka:2015iiw,
    author = "Nosaka, Tomoki",
    title = "{Instanton effects in ABJM theory with general R-charge assignments}",
    eprint = "1512.02862",
    archivePrefix = "arXiv",
    primaryClass = "hep-th",
    reportNumber = "YITP-15-105",
    doi = "10.1007/JHEP03(2016)059",
    journal = "JHEP",
    volume = "03",
    pages = "059",
    year = "2016"
}

@article{Chester:2021gdw,
    author = "Chester, Shai M. and Kalloor, Rohit R. and Sharon, Adar",
    title = "{Squashing, Mass, and Holography for 3d Sphere Free Energy}",
    eprint = "2102.05643",
    archivePrefix = "arXiv",
    primaryClass = "hep-th",
    doi = "10.1007/JHEP04(2021)244",
    journal = "JHEP",
    volume = "04",
    pages = "244",
    year = "2021"
}

@article{Freedman:2013oja,
    author = "Freedman, Daniel Z. and Pufu, Silviu S.",
    title = "{The holography of $F$-maximization}",
    eprint = "1302.7310",
    archivePrefix = "arXiv",
    primaryClass = "hep-th",
    reportNumber = "SU-ITP-13-01, MIT-CTP-4443",
    doi = "10.1007/JHEP03(2014)135",
    journal = "JHEP",
    volume = "03",
    pages = "135",
    year = "2014"
}

@article{Martelli:2011fu,
    author = "Martelli, Dario and Passias, Achilleas and Sparks, James",
    title = "{The gravity dual of supersymmetric gauge theories on a squashed three-sphere}",
    eprint = "1110.6400",
    archivePrefix = "arXiv",
    primaryClass = "hep-th",
    doi = "10.1016/j.nuclphysb.2012.07.019",
    journal = "Nucl. Phys. B",
    volume = "864",
    pages = "840--868",
    year = "2012"
}

@article{Kurlyand:2026cyz,
    author = "Kurlyand, Stefan A. and Tseytlin, Arkady A.",
    title = "{1-loop partition function of 11d supergravity on $\mathrm{AdS}_4\times S^7/\mathbb{Z}_k$: matching $A(k)$ in ABJM free energy}",
    eprint = "2611.15510",
    archivePrefix = "arXiv",
    primaryClass = "hep-th",
    reportNumber = "Imperial--TP--2026--SK--04",
    month = "11",
    year = "2026"
}

@article{Pestun:2007rz,
    author = "Pestun, Vasily",
    title = "{Localization of gauge theory on a four-sphere and supersymmetric Wilson loops}",
    eprint = "0712.2824",
    archivePrefix = "arXiv",
    primaryClass = "hep-th",
    reportNumber = "ITEP-TH-41-07, PUTP-2248",
    doi = "10.1007/s00220-012-1485-0",
    journal = "Commun. Math. Phys.",
    volume = "313",
    pages = "71--129",
    year = "2012"
}

@article{Pestun:2016zxk,
    author = "Pestun, Vasily and others",
    title = "{Localization techniques in quantum field theories}",
    eprint = "1608.02952",
    archivePrefix = "arXiv",
    primaryClass = "hep-th",
    doi = "10.1088/1751-8121/aa63c1",
    journal = "J. Phys. A",
    volume = "50",
    number = "44",
    pages = "440301",
    year = "2017"
}

@article{Binder:2019jwn,
    author = "Binder, Damon J. and Chester, Shai M. and Pufu, Silviu S. and Wang, Yifan",
    title = "{$ \mathcal{N} $ = 4 Super-Yang-Mills correlators at strong coupling from string theory and localization}",
    eprint = "1902.06263",
    archivePrefix = "arXiv",
    primaryClass = "hep-th",
    reportNumber = "PUPT-2582",
    doi = "10.1007/JHEP12(2019)119",
    journal = "JHEP",
    volume = "12",
    pages = "119",
    year = "2019"
}

@article{Chester:2019jas,
    author = "Chester, Shai M. and Green, Michael B. and Pufu, Silviu S. and Wang, Yifan and Wen, Congkao",
    title = "{Modular invariance in superstring theory from $ \mathcal{N} $ = 4 super-Yang-Mills}",
    eprint = "1912.13365",
    archivePrefix = "arXiv",
    primaryClass = "hep-th",
    reportNumber = "PUPT-2607, QMUL-PH-19-37",
    doi = "10.1007/JHEP11(2020)016",
    journal = "JHEP",
    volume = "11",
    pages = "016",
    year = "2020"
}

@article{Binder:2018yvd,
    author = "Binder, Damon J. and Chester, Shai M. and Pufu, Silviu S.",
    title = "{Absence of $D^4 R^4$ in M-Theory From ABJM}",
    eprint = "1808.10554",
    archivePrefix = "arXiv",
    primaryClass = "hep-th",
    reportNumber = "PUPT-2570",
    doi = "10.1007/JHEP04(2020)052",
    journal = "JHEP",
    volume = "04",
    pages = "052",
    year = "2020"
}

@article{Chester:2020dja,
    author = "Chester, Shai M. and Pufu, Silviu S.",
    title = "{Far beyond the planar limit in strongly-coupled $ \mathcal{N} $ = 4 SYM}",
    eprint = "2003.08412",
    archivePrefix = "arXiv",
    primaryClass = "hep-th",
    reportNumber = "PUPT-2616",
    doi = "10.1007/JHEP01(2021)103",
    journal = "JHEP",
    volume = "01",
    pages = "103",
    year = "2021"
}

@article{DiPietro:2014bca,
    author = "Di Pietro, Lorenzo and Komargodski, Zohar",
    title = "{Cardy formulae for SUSY theories in $d =$ 4 and $d =$ 6}",
    eprint = "1407.6061",
    archivePrefix = "arXiv",
    primaryClass = "hep-th",
    doi = "10.1007/JHEP12(2014)031",
    journal = "JHEP",
    volume = "12",
    pages = "031",
    year = "2014"
}

@article{Benjamin:2023qsc,
    author = "Benjamin, Nathan and Lee, Jaeha and Ooguri, Hirosi and Simmons-Duffin, David",
    title = "{Universal asymptotics for high energy CFT data}",
    eprint = "2306.08031",
    archivePrefix = "arXiv",
    primaryClass = "hep-th",
    reportNumber = "CALT-TH 2023-014, IPMU 23-0020",
    doi = "10.1007/JHEP03(2024)115",
    journal = "JHEP",
    volume = "03",
    pages = "115",
    year = "2024"
}

@article{Green:2006gt,
    author = "Green, Michael B. and Russo, Jorge G. and Vanhove, Pierre",
    title = "{Non-renormalisation conditions in type II string theory and maximal supergravity}",
    eprint = "hep-th/0610299",
    archivePrefix = "arXiv",
    reportNumber = "DAMTP-2006-102, SPHT06-127, UB-ECM-PF-06-30",
    doi = "10.1088/1126-6708/2007/02/099",
    journal = "JHEP",
    volume = "02",
    pages = "099",
    year = "2007"
}

@unpublished{Herderschee:2026IIR4,
    author = "Herderschee, Aidan and Ivo, Victor and Lin, Henry W. and Wang, Yifan",
    note = "in preparation",
    year = "2026"
}

@article{Bhattacharyya:2012ye,
    author = "Bhattacharyya, Sayantani and Grassi, Alba and Marino, Marcos and Sen, Ashoke",
    title = "{A One-Loop Test of Quantum Supergravity}",
    eprint = "1210.6057",
    archivePrefix = "arXiv",
    primaryClass = "hep-th",
    doi = "10.1088/0264-9381/31/1/015012",
    journal = "Class. Quant. Grav.",
    volume = "31",
    pages = "015012",
    year = "2014"
}

@article{Vassilevich:2003xt,
    author = "Vassilevich, D. V.",
    title = "{Heat kernel expansion: User's manual}",
    eprint = "hep-th/0306138",
    archivePrefix = "arXiv",
    doi = "10.1016/j.physrep.2003.09.002",
    journal = "Phys. Rept.",
    volume = "388",
    pages = "279--360",
    year = "2003"
}

@article{Sen:2012kpz,
    author = "Sen, Ashoke",
    title = "{Logarithmic Corrections to N=2 Black Hole Entropy: An Infrared Window into the Microstates}",
    eprint = "1108.3842",
    archivePrefix = "arXiv",
    primaryClass = "hep-th",
    doi = "10.1007/s10714-012-1336-5",
    journal = "Gen. Rel. Grav.",
    volume = "44",
    number = "5",
    pages = "1207--1266",
    year = "2012"
}

@article{Siegel:1980jj,
    author = "Siegel, W.",
    title = "{Hidden Ghosts}",
    reportNumber = "Print-80-0192 (IAS,PRINCETON)",
    doi = "10.1016/0370-2693(80)90119-7",
    journal = "Phys. Lett. B",
    volume = "93",
    pages = "170--172",
    year = "1980"
}

@article{Nielsen:1978mp,
    author = "Nielsen, N. K.",
    title = "{Ghost Counting in Supergravity}",
    reportNumber = "NORDITA-78/11",
    doi = "10.1016/0550-3213(78)90009-3",
    journal = "Nucl. Phys. B",
    volume = "140",
    pages = "499--509",
    year = "1978"
}

@article{Kallosh:1978de,
    author = "Kallosh, R. E.",
    title = "{Modified Feynman Rules in Supergravity}",
    doi = "10.1016/0550-3213(78)90340-1",
    journal = "Nucl. Phys. B",
    volume = "141",
    pages = "141--152",
    year = "1978"
}

@article{Visser:1992pz,
    author = "Visser, Matt",
    title = "{van Vleck determinants: Geodesic focusing and defocusing in Lorentzian space-times}",
    eprint = "hep-th/9303020",
    archivePrefix = "arXiv",
    reportNumber = "WASH-U-HEP-92-76",
    doi = "10.1103/PhysRevD.47.2395",
    journal = "Phys. Rev. D",
    volume = "47",
    pages = "2395--2402",
    year = "1993"
}

@article{tHooft:1974toh,
    author = "'t Hooft, Gerard and Veltman, M. J. G.",
    title = "{One-loop divergencies in the theory of gravitation}",
    doi = "10.1142/9789814539395_0001",
    journal = "Ann. Inst. H. Poincare Phys. Theor. A",
    volume = "20",
    number = "1",
    pages = "69--94",
    year = "1974"
}

@article{Christensen:1979iy,
    author = "Christensen, S. M. and Duff, M. J.",
    title = "{Quantizing Gravity with a Cosmological Constant}",
    reportNumber = "NSF-ITP-79-01",
    doi = "10.1016/0550-3213(80)90423-X",
    journal = "Nucl. Phys. B",
    volume = "170",
    pages = "480--506",
    year = "1980"
}

@article{Camporesi:1990wm,
    author = "Camporesi, R.",
    title = "{Harmonic analysis and propagators on homogeneous spaces}",
    doi = "10.1016/0370-1573(90)90120-Q",
    journal = "Phys. Rept.",
    volume = "196",
    pages = "1--134",
    year = "1990"
}

@article{Liu:2016dau,
    author = "Liu, James T. and Zhao, Wenli",
    title = "{One-loop supergravity on $\mathrm{AdS}_4\times S^7/\mathbb{Z}_k$ and comparison with ABJM theory}",
    eprint = "1609.02558",
    archivePrefix = "arXiv",
    primaryClass = "hep-th",
    reportNumber = "MCTP-16-21",
    doi = "10.1007/JHEP11(2016)099",
    journal = "JHEP",
    volume = "11",
    pages = "099",
    year = "2016"
}

\end{document}